\documentclass[12pt]{article}
\usepackage{amsmath,amsthm, amsfonts, amssymb, amsxtra,amsopn}
\usepackage{graphicx}
\usepackage{multirow}
\usepackage{tikz}
\usepackage{cmap}
\usepackage{pgfplots}
\pgfplotsset{compat=1.7}
\usepackage{pgfplotstable}
\PassOptionsToPackage{hyphens}{url}
\usepackage{hyperref}
\hypersetup{colorlinks=true,linkcolor=black,citecolor=black,urlcolor=black,filecolor=black}
\hypersetup{pdfpagemode=UseNone,pdfstartview=}

\graphicspath{{./images/}}

\advance\oddsidemargin by -0.6in
\advance\textwidth by 1.2in

\advance\topmargin by -0.36in
\advance\textheight by 0.72in

\def\eref#1{(\ref{#1})}

\def\subsubsection#1{\bigskip {\noindent\bf #1} ---}

\def\e{\varepsilon}
\def\E{\mbox{E}}
\def\C{\mbox{C}}
\def\G{\mbox{G}}
\def\J{\mbox{J}}

\def\z{\phantom{0}}

\def\O{{\cal O}}

\long\def\symbolfootnotetext[#1]#2{\begingroup%
\def\thefootnote{\fnsymbol{footnote}}\footnotetext[#1]{#2}\endgroup}

\title{Malware Detection Using Dynamic Birthmarks}

\author{Swapna Vemparala\footnotemark[1]\ \ \ 
Fabio Di Troia\footnotemark[2]\ \ 
Visaggio Aaron Corrado\footnotemark[2]\\
Thomas H.~Austin\footnotemark[1]\ \ 
Mark Stamp\footnotemark[1]\,\,\,\footnotemark[3]}

\begin{document}

\symbolfootnotetext[1]{Department of Computer Science, San Jose State University}
\symbolfootnotetext[2]{Department of Engineering, Universit\`{a} degli Studi del Sannio}
\symbolfootnotetext[3]{stamp$@$cs.sjsu.edu}

\maketitle

%
%
\abstract

In this paper, we explore the effectiveness of dynamic analysis techniques
for identifying malware, using Hidden Markov Models (HMMs) 
and Profile Hidden Markov Models (PHMMs),
both trained on sequences of API calls.
We contrast our results to static analysis using HMMs trained on sequences of opcodes,
and show that dynamic analysis achieves significantly stronger results in many cases.
Furthermore, in contrasting our two dynamic analysis techniques,
we find that using PHMMs consistently outperforms our analysis based on HMMs.

%
%
\section{Introduction}

News stories abound about cyber attacks relating to malware.
In 2014, Twitter was at the receiving end of a major cyber attack.
According to news reports, 250,000 users' email addresses, user names, and passwords were compromised.
Twitter was able to detect the attack by identifying the abnormal patterns in which data was accessed~\cite{twitAttack}.

Target fell victim to a major security breach during the 2013 holiday season,
where credit and debit card details of more than a million customers were compromised.
The information was stolen by hacking the credit card swipe systems at Target stores~\cite{targetAttack}.
This one attack drove down quarterly revenues of Target by tens of millions of dollars~\cite{targetAttackWashPost}.

In today's world of malware and cyber threats,
it has become critical to develop techniques that quickly identify malware.
In this paper, we look at different malware detection techniques and compare their results.

We use the concept of software birthmarks for malware detection.
Software birthmarks are inherent characteristics that can be 
used to identify particular software~\cite{myles,tamada}.
The goal is to obtain a unique identifier for each executable.
We can then use these birthmarks
to test the similarity between two executables.
If the birthmarks of the two files are sufficiently similar,
then we assume that one software is closely related to the other.
This strategy has been the basis of a variety of techniques for identifying metamorphic malware
with statistical approaches~\cite{fukuda,kazi,myles,rana,tamada,tamada2,Wang,zhou}.

Software birthmarks can be either static or dynamic~\cite{fukuda}.
Static birthmarks are characteristics that can be extracted from a program without executing it~\cite{zhou}.
For example, a static birthmark can be based on an extracted opcode sequence.
In contrast, dynamic birthmarks are obtained from a program when it is executed~\cite{kazi,tamada,zhou}.
An example of a dynamic birthmark is the sequence of API calls that occur when a program is executed~\cite{tamada2}.

Previous work explored static birthmarks for malware detection~\cite{tamada2};
this paper instead considers dynamic birthmarks. Specifically,
we use Hidden Markov Models (HMMs) and Profile Hidden Markov Models (PHMMs) 
as the bases for our detection techniques.
Both HMM-based analysis~\cite{wongStamp} and PHMM-based analysis~\cite{bib28}
have been previously applied to the malware detection problem.

To compare the effectiveness of our dynamic analysis techniques,
we compare static and dynamic analysis results on substantial malware datasets
using Receiving Operating Characteristic (ROC) analysis~\cite{Bradley}.
We show that significantly stronger malware detection results can be obtained
using dynamic analysis of software birthmarks, as compared to static analysis.
Stated another way, our results suggest that it is more beneficial
to consider the API calls used during program execution
than to statically analyze the opcodes in a program.
Our work also demonstrates that PHMMs are more effective than HMMs in malware analysis
when considering this particular type of dynamic birthmark.
In our tests, properly tuned PHMMs built from dynamic birthmarks always outperform
HMMs built from the same birthmarks.


The remainder of this paper is organized as follows.
In Section~\ref{sec:malBack}, we provide background information on malware and detection techniques.
Sections~\ref{sec:hmmOverview} and~\ref{sec:phmmOverview} respectively
cover Hidden Markov Models and Profile Hidden Markov Models.
These are the machine learning techniques used in this paper.
Section~\ref{sect:relatedWork} gives a brief survey of related work involving
these machine learning techniques.
In Section~\ref{chap:Implementation},
we discuss implementation details.
Here, we discuss in detail the tools used for extracting the data,
the dataset used, and the application of the machine learning techniques used.
Experiments and discussion of the results is given in Section~\ref{chap:Experiments and Results}.
Section~\ref{chap:Conclusion and Future work} concludes and we briefly consider future work.


\section{Malware\label{sec:malBack}}

In this section, we discuss relevant topics related to malware. First, we 
briefly consider different types of malware. Then we discuss methods
used to obfuscate malware, which is followed by an overview of some advanced
anti-virus techniques. We conclude with a discussion of software
birthmarks, in the context of malware analysis.

\subsection{Types of Malware}\label{sec:malwareTypes}

Malware is a generic term that encompasses viruses, worms, backdoors, and Trojans.
In this section, we discuss the different types of malware in detail.
Note that here we are primarily concerned
with the transmission technique of the malware.
Section~\ref{sec:detectionEvasion} discusses different obfuscation
techniques that might be employed by malware writers,
and could be applied to any of the forms of malware listed here.


``Virus'' is perhaps the most overloaded term used for 
malware---in the vernacular, virus is a synonym for malware.
Technical usage is also sometimes ambiguous.
The term virus is sometimes used to refer to self-replicating malware~\cite{szorBook},
and by this definition includes worms, for instance.
In this paper we us a stricter definition:
a virus is a type of malware that replicates itself by infecting other files,
that is, a virus is parasitic.
This definition seems to be somewhat more common,
and is the definition used, for example, by Cohen~\cite{szorBook}.

A germ file (the initial malicious program) infects other, benign files
and inserts the virus code into the execution of the benign program.
When these infected program are executed,
the virus can insert itself into another generation of victim programs.
For example,
Word macro viruses execute when the user opens an infected Word document,
executing their payload and infecting other Word documents~\cite{aycock}.

Stuxnet, created in 2010, is a famous examples of a virus.
It was primarily developed to target the industrial control systems used in power plants.
This malware attacked the programmable logic controllers present in the control systems~\cite{stuxnet}.

In subsequent sections, we often use the term virus 
in its vernacular sense. That is, we use the terms ``virus''
and ``malware'' interchangeably.


A Trojan is malware that looks innocuous to the user, 
but performs malicious activities in the background.
Unlike a virus, a Trojan is a standalone program.
Trojans are generally not self-replicating in nature and depend
on users for their propagation.
As an example, an attacker can create a 
login program that prompts users to enter their usernames and passwords.
If the user enters his or her credentials,
the program steals the users login credentials and displays an invalid login message
and then runs the actual login program~\cite{aycock}.
Trojans can be very harmful,
as the user is completely unaware of the malicious activities that take place in the background.
Trojans constitute a major portion of the today's malware activity~\cite{bib22}.


Worms are defined primarily by their method of propagation;
they self-replicate to different systems through the network~\cite{szorBook}.
Unlike viruses, worms are not parasitic in behavior,
meaning that they do not alter the instructions of other programs.
Worms have accounted for some well-known malware exploits,
including the Morris worm, Code Red, and SQL Slammer~\cite{bib8}.


A backdoor is used to bypass the normal security authentication processes,
and could be installed as the result of a worm or virus.
After gaining access, it takes control
and can remotely monitor the system while stealing personal information~\cite{aycock}.
An example for a backdoor malware is Back Orifice.
On gaining control of the infected system,
this malware can download, upload, or delete files,
lock or shutdown the computer, or even take control of the keyboard and mouse~\cite{bib17}.

\subsection{Malware Obfuscation Strategies}\label{sec:detectionEvasion}

Signature-based detection is the most common strategy used by anti-virus products.
This approach involves scanning through the files in a system in search of specific bit patterns.
An anti-virus product has a database of patterns or signatures,
and if a match is found, then that file is likely to be malware.
It is essential that the signature database be updated frequently;
otherwise new malware will go undetected.

Signature detection has long been the main approach used to identify malware,
largely due to its efficiency and low false positive rate.
Therefore, virus authors have worked to change the signature of their malware
while still providing the same functionality.


Encryption was one of the first obfuscation strategies used by malware writers,
and was seen, for example, in the DOS virus Cascade~\cite{virginia}.
By encrypting the body of the malware with a different key,
the signature of the payload will be different.
During execution, the encrypted code can be decrypted and then run normally.

While encrypting the payload of the virus results in a different signature,
the virus must decrypt itself before executing its payload.
Therefore, the decryption code may be identified by signature detection.

Typically, simple encryption techniques are used, since there is little benefit to the virus user
of a strong encryption mechanism.
The encryption can be done by means of a static or a variable key,
a simple bitwise rotation, or even by incrementing or decrementing the bits in the virus body~\cite{aycock}.
In choosing a decryption algorithm, the virus writer might attempt 
to use an algorithm that can evade signature detection;
more advanced variants of encrypted viruses focus on exactly this problem.


Polymorphic viruses~\cite{polymorphic}
are a variant of encrypted viruses where the decryption code changes in each generation,
unlike the case of encrypted viruses, whose decryption code remains constant.
When executed, polymorphic viruses decrypt and execute their payload
in the same manner as with encrypted viruses.
Examples of polymorphic viruses include
Tequila and Maltese~\cite{polymorphic}.

Polymorphic viruses work by mutating the code of the decryption routine,
producing equivalent decryption routines with different instructions,
and therefore different signatures.
(We discuss the code mutation techniques in more detail when we review metamorphic viruses, below.)
With a wide array of possible versions of the decryption code,
signature detection is much more difficult, if not impossible~\cite{aycock}.

Code emulation is effective for detecting polymorphic malware.
The suspect file is executed for some period of time within a virtual machine
until the virus decrypts itself~\cite{Metamorphic}.
At that point, standard signature detection techniques can be applied to the file in question.


Metamorphic viruses take polymorphic viruses to the logical extreme.
Rather than relying on encryption, morphing techniques are used on the body of the virus;
for this reason, metamorphic viruses are sometimes described as ``body polymorphic''~\cite{szorBook}.

A metamorphic virus makes use of a mutation engine in order to morph its body at each generation,
but the essentially functionality remains unchanged~\cite{aycock}.
Metamorphic viruses can make use of different methods to morph the internal structure.
A few of these methods are discussed below.

\newcommand{\myitem}[1]{\item {\em #1.}}
\begin{enumerate}

\myitem {Dead Code Insertion}
This technique is used in many metamorphic viruses.
The idea here is to insert dead code in such a way that it does not
alter the functionality of the original code.
The Win32/Evol virus 
includes a metamorphic engine that inserts garbage code~\cite{Metamorphic}.

\myitem {Instruction Reordering}
This strategy changes the order of instructions, thereby breaking signatures.
As long as no code dependency is broken, instruction reordering does not affect the functionality of the virus.
Instruction reordering can be used to generate a large number of morphed copies.

\myitem {Instruction Substitution}
This approach involves replacing a group of instructions with equivalent instructions.
The functionality of the malware remains unchanged,
but the morphed versions can evade signature detection~\cite{aycock}.

\myitem {Register Swapping}
In this technique, the virus body is morphed by swapping operand registers with different registers.
Though the virus body is mutated,
this technique does not change the opcode sequence.
The W95/Regswap virus uses register swapping~\cite{Metamorphic}.

\end{enumerate}

\subsection{Anti-Virus Techniques}

In the face of increasingly effective techniques for evading signature detection,
advanced detection strategies have been developed.
In this section, we review some of these approaches.


Unlike signature based detection, static heuristics analysis does not identify malware by scanning
for specific signatures.
Instead, it analyses the behavior, structure, and other attributes for virus-like qualities~\cite{aycock}.
This technique can be used to identify zero-day malware as well as existing malware.
This method identifies the probability of a file being infected by performing an analysis
of the instructions, the logic of the program, the data used, and the overall structure of the binary file it scans.
While static heuristics analysis has the ability to detect malware even before they can infect the computer,
they tend to have a high false positive rate~\cite{heuristic}.


A behavior blocker executes the program and closely monitors it for suspicious activities.
If any suspicious activity is encountered, the behavior blocker stops the program~\cite{aycock}.
Behavior blockers make use of dynamic signatures,
which are created by collecting during program execution.

In general, dynamic detection techniques 
only consider the instructions that have actually been executed.
Dead code insertion, for example,
has no impact on dynamic analysis. Consequently, we
can view dynamic analysis as stripping away one layer of
obfuscation. However, the tradeoff is that dynamic analysis
is based only on the parts of the code that execute.

%


Unlike behavior blockers,
emulators do not run the malware directly on the system.
Instead, they execute the code in an emulated environment.
Emulators make use of dynamic heuristics,
which collect information about the malware that is being executed in the emulated environment.
Given the fact that dynamic heuristics closely monitor the operating system,
they have a significant advantage over static heuristics in detecting malware that targets operating systems.
On the negative side, the CPU emulation process is much slower compared to signature scanning,
and hence the dynamic heuristics is slower than static heuristics.
Also some viruses attempt to hide their infection logic from the emulator
by performing the infection only under specific conditions~\cite{heuristic}.

\subsection{Software Birthmarks}

Software birthmarks are characteristics that are derived from a particular software.
There are two types of birthmarks---static and dynamic.
Static birthmarks are characteristics that can be extracted from the program itself without executing it.
For example, opcodes sequences can be considered a static birthmark for a program.
Dynamic birthmarks are characteristics that are obtained from a program when it is executed.
For example, API calls that are recorded when the program is executed could serve
as a dynamic birthmark.
Next, we briefly discuss examples of analysis techniques
that use static and dynamic birthmarks for malware detection


As previously mentioned, static analysis involves analyzing the program without executing it.
In~\cite{rana}, static analysis is performed by extracting opcode sequences with the help of a disassembler.
The key advantage of static analysis is that it is relatively faster and efficient.
Compared to dynamic analysis techniques, it is also safer since it does not involve execution of the malware.
The drawback of a static approach is that it may be defeated by code obfuscation techniques~\cite{aycock}.

In~\cite{myles}, the authors propose a static $k$-gram based birthmark.
This birthmark is computed by extracting opcode sequences of length~$k$.
The effectiveness of this birthmark is tested on randomly selected applications
and on programs that are semantically different, but which accomplish the same task.
The authors identify an optimum~$k$ value at which the credibility and resilience of the birthmark is maximized.

Dynamic analysis is performed to extract dynamic software birthmarks.
Dynamic birthmarks, such as API calls obtained at runtime, can be difficult for
malware writers to defeated~\cite{tamada}, since
dynamic birthmarks are resilient to code obfuscation when compared to static birthmarks.
The tradeoff is that the cost of dynamic analysis is generally higher
and there is risk involved in executing the malware.

In~\cite{K-gram}, the authors construct a $k$-gram based birthmark by traversing a window of length~$k$
over an opcode sequence in the case of the static analysis and over an executable trace for the dynamic analysis case.
The authors evaluated the strength of the dynamic birthmark technique by comparing the results obtained
from static and dynamic cases when code obfuscation is applied.
They show that the dynamic $k$-gram birthmark is much more robust than the static case.

Anderson~\cite{Anderson} discusses a graph based technique in which a graph is constructed
by collecting dynamic instruction traces.
The constructed graphs are Markov chains consisting of instructions and transition probabilities.
A similarity matrix is created using the graph kernels for different instruction traces.
A Support Vector Machine is applied to the similarity matrix to perform the malware classification.
The author of~\cite{Anderson} shows that this technique is an improvement over 
previous work.

Tamada et al.~\cite{tamada}
use both the API call sequences and the API call frequencies that are collected during runtime as dynamic birthmarks.
The authors leverage the fact that different programs using the same APIs 
do not have the same order of API calls upon execution.

Fukuda et al.~\cite{fukuda} consider the operand stack runtime behavior of a Java Virtual Machine.
This is a dynamic birthmark, due to its unique nature at runtime.
This birthmark exhibits high tolerance to code changes and a good ability to discern programs
that could be a pirated version of another.

It is evident from the works above that dynamic birthmarks have 
some inherent advantages as compared to their static counterparts.
Our work consists of developing and analyzing malware detection strategies
based on dynamic birthmarks and using Hidden Markov Models and Profile Hidden Markov Models
for classification.

\section{Hidden Markov Models}\label{sec:hmmOverview}

In this section, we discuss Hidden Markov Models (HMMs) in some detail.
HMMs have proven useful in a wide array of fields,
including speech recognition~\cite{Rabiner}, 
malware detection~\cite{wongStamp},
and software piracy analysis~\cite{rana}.
In this paper, we apply HMMs
to the problem of malware detection based on software birthmarks.

A Markov process of order one is a memoryless process where
the current state depends only on the prvious state.
In a Hidden Markov Model, the state is hidden, in the sense that it cannot be directly observed.
However, we do have observations that depend (probabilistically) on the hidden states.

\begin{figure*}[htb]
  \begin{center}
    \begin{tikzpicture}[thick,scale=1.0]
    
    \draw[thick,color=blue] (0,0) rectangle (1,1);
    \draw[thick,color=blue] (2.5,0) rectangle (3.5,1);
    \draw[thick,color=blue] (5,0) rectangle (6,1);
    \draw[thick,color=blue] (10,0) rectangle (11,1);

    \draw[thick,color=green] (0.5,4.5) circle (0.575);
    \draw[thick,color=green] (3,4.5) circle (0.575);
    \draw[thick,color=green] (5.5,4.5) circle (0.575);
    \draw[thick,color=green] (10.5,4.5) circle (0.575);
    
    \node at (0.5,0.5){$\O_0$};
    \node at (3,0.5){$\O_1$};
    \node at (5.5,0.5){$\O_2$};
    \node at (8,0.5){$\cdots$};
    \node at (10.5,0.5){$\O_{T-1}$};

    \node at (0.5,4.5){$X_0$};
    \node at (3,4.5){$X_1$};
    \node at (5.5,4.5){$X_2$};
    \node at (8,4.5){$\cdots$};
    \node at (10.5,4.5){$X_{T-1}$};
       
    \node at (1.7,4.8){$A$};
    \node at (4.2,4.8){$A$};
    \node at (6.7,4.8){$A$};
    \node at (9.2,4.8){$A$};
    
    \node at (0.2,2.1){$B$};
    \node at (2.7,2.1){$B$};
    \node at (5.2,2.1){$B$};
    \node at (10.2,2.1){$B$};
    
     \draw[thick,color=black,->] (1.075,4.5) -- (2.425,4.5);
     \draw[thick,color=black,->] (3.575,4.5) -- (4.925,4.5);
     \draw[thick,color=black,->] (6.075,4.5) -- (7.425,4.5);
     \draw[thick,color=black,->] (8.575,4.5) -- (9.925,4.5);

     \draw[thick,color=black,->] (0.5,3.925) -- (0.5,1);
     \draw[thick,color=black,->] (3.0,3.925) -- (3.0,1);
     \draw[thick,color=black,->] (5.5,3.925) -- (5.5,1);
     \draw[thick,color=black,->] (10.5,3.925) -- (10.5,1);

    \draw[thick,dashed,color=red] (-0.3,3) -- (11.2,3);
   
    \end{tikzpicture}
  \end{center}
  \caption{Hidden Markov Model.\label{fig:hmm}} 
\end{figure*}
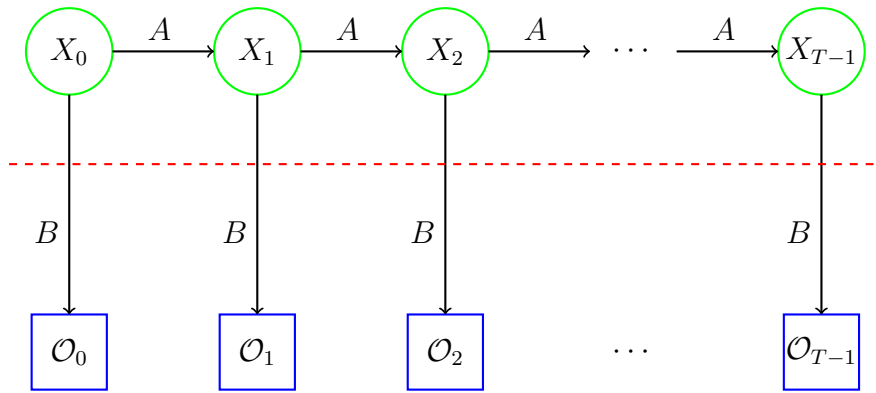


Each state of an HMM
has an associated probability distribution over the set of possible observations.
Since we have a Markov process, the transition probabilities between the states are fixed.
For our research we train an HMM using observation sequences from malware samples
belonging to a given family.
After training, we can score any given observation sequence against the trained HMM to determine
the likelihood of the sequence belonging to the same family that was used to train the HMM.
A high probability indicates the scored sample closely matches
the sequence used to train the HMM~\cite{bib18}.

We use the standard notation in Table~\ref{tab:HMMnotation}
in our discussion of HMMs.
\begin{table}[htb]
  \begin{center}
  \caption{HMM Notation\label{tab:HMMnotation}}
  \begin{tabular}{cl} \hline\hline
    notation & \hspace*{0.5in}explanation\\ \hline
    $T$ & length of the observation sequence\\
    $N$ & number of states in the model\\
    $M$ & number of observation symbols\\
    $Q$ & distinct states, $\{q_0,q_1,\ldots,q_{N-1}\}$\\
    $V$ & set of possible observations, $\{0,1,\ldots,M-1\}$\\
    $A$ & state transition probabilities\\
    $B$ & observation probability matrix\\
    $\pi$ & initial state distribution\\
    $\O$ & observation sequence $(\O_0,\O_1,\ldots,\O_{T-1})$\\ \hline\hline
  \end{tabular}
\end{center}
\end{table}
The matrices~$A$, $B$, and~$\pi$ provide all relevant information about the
model and hence we denote an HMM as~$\lambda = (A, B, \pi)$.

Figure~\ref{fig:hmm} provides a generic illustration of an HMM.
In Figure~\ref{fig:hmm}, the~$X_i$ represent the hidden states.
The (hidden) Markov process is determined by 
the~$A$ matrix and the present state.
The observations~$\O_i$ are known and are related to the hidden states by the~$B$ matrix.
Hence, we can indirectly obtain information about the hidden states
from the observation sequence and the~$B$ matrix.
The matrices~$A$, $B$ and~$\pi$ are all row stochastic, that is,
each row satisfies the conditions of a discrete probability distribution.

HMMs are useful primarily because there exist efficient algorithms
to solve each of the following three problems~\cite{bib18}.

\subsubsection{Problem 1} Given a model~$\lambda = (A, B, \pi)$ and an observation sequence~$\O$,
determine~$P(\O\,|\,\lambda)$. That is, we can score
a given observation sequence against a given model.

\subsubsection{Problem 2} Given a model~$\lambda = (A, B, \pi)$ and an observation sequence~$\O$,
determine an optimal state sequence for the underlying Markov model.
That is, we can ``uncover'' the most likely hidden state sequence.

\subsubsection{Problem 3} Given a observation sequence~$\O$ we can 
generate a model~$\lambda = (A, B, \pi)$ such that~$P(\O\,|\,\lambda)$ is maximized.
That is, we can train a model to fit a given observations sequence.

\bigskip

Next, we discuss the solution to each of these three problems. For additional
details, see the tutorial~\cite{bib18} 
or Rabiner's classic introduction~\cite{Rabiner}.

\subsubsection{Solution to Problem 1}
Let~$\lambda=(A,B,\pi)$ be a given model
and let $\O=(O_0,\O_1,\ldots,\O_{T-1})$ be a series of observations.
We want to find~$P(\O\,|\,\lambda)$. 

Let $X=(x_0,x_1,\ldots,x_{T-1})$ 
be a state sequence. Then by the definition of~$B$ we have
$$
  P(\O\,|\,X,\lambda) = b_{x_0}(\O_0)b_{x_1}(\O_1)\cdots 
      b_{x_{T-1}}(\O_{T-1})
$$
and by the definition of~$\pi$ and~$A$ it follows that
$$
  P(X\,|\,\lambda) = \pi_{x_0}a_{x_0,x_1}a_{x_1,x_2}\cdots a_{x_{T-2},x_{T-1}} .
$$
Since
$$
  P(\O,X\,|\,\lambda) = \frac{P(\O\cap X\cap \lambda)}{P(\lambda)}
$$
and
\begin{align*}
  P(\O\,|\,X,\lambda)P(X\,|\,\lambda) 
    &= \frac{P(\O\cap X\cap \lambda)}{P(X\cap\lambda)} \cdot 
      \frac{P(X\cap\lambda)}{P(\lambda)} \\
    &= \frac{P(\O\cap X\cap \lambda)}{P(\lambda)}
\end{align*}
we have
$$
  P(\O,X\,|\,\lambda) = P(\O\,|\,X,\lambda)P(X\,|\,\lambda) .
$$
By summing over all possible state sequences we obtain
\begin{align}\label{eq:HMMxo}
  P(\O\,|\,\lambda) &= \sum_{X} P(\O,X\,|\,\lambda) \nonumber\\
                    &= \sum_{X}
                        P(\O\,|\,X,\lambda)P(X\,|\,\lambda)\\
                    &= \sum_{X} 
                        \pi_{x_0}b_{x_0}(\O_0)a_{x_0,x_1}b_{x_1}(\O_1)\cdots \\ \nonumber
                    &\hspace*{0.5in} \cdots a_{x_{T-2},x_{T-1}}b_{x_{T-1}}(\O_{T-1}) . \nonumber
\end{align}
However, this direct computation is generally computationally infeasible.

To compute~$P(\O\,|\,\lambda)$ efficiently,
the so-called forward algorithm is used. 
For $t=0,1,\ldots,T-1$ and $i=0,1,\ldots,N-1$,
define
\begin{equation}\label{eq:alpha}
  \alpha_t(i) = P(\O_0,\O_1,\ldots,\O_t,x_t=q_i\,|\,\lambda) .
\end{equation}
Then~$\alpha_t(i)$ is the probability of the partial observation
sequence up to time~$t$, where the underlying Markov process is in 
state~$q_i$ at time~$t$. 

The~$\alpha_t(i)$ can be computed recursively as follows.
\begin{enumerate}
\item Let $\alpha_0(i)=\pi_i b_i(\O_0)$, for $i=0,1,\ldots,N-1$
\item For $t=1,2,\ldots,T-1$ and $i=0,1,\ldots,N-1$, compute
$$
  \alpha_{t}(i) = \Biggl(\sum_{j=0}^{N-1}\alpha_{t-1}(j)a_{ji}\Biggr) 
                   b_i(\O_{t})
$$
\item Then from~\eref{eq:alpha}, we have
$$
  P(\O\,|\,\lambda) = \sum_{i=0}^{N-1}\alpha_{T-1}(i) .
$$
\end{enumerate}
The forward algorithm only requires about~$N^2T$ multiplications, 
as opposed to more than~$2TN^T$ for the na\"{i}ve approach.

\subsubsection{Solution to Problem 2}
Given the model~$\lambda=(A,B,\pi)$ and
a sequence of observations~$\O$, we can find the 
most likely state sequence. There are
different possible interpretations of ``most likely.'' For
HMMs, ``most likely'' means that we maximize the expected number of correct
states. 

In analogy to the forward algorithm, 
we define the backward algorithm as follows.
Let
$$
  \beta_t(i) = P(\O_{t+1},\O_{t+2},\ldots,\O_{T-1}\,|\,x_t=q_i,\lambda) 
$$
for~$t=0,1,\ldots,T-1$ and~$i=0,1,\ldots,N-1$.
Then the~$\beta_t(i)$ can be computed recursively
(and efficiently) using the following algorithm.
\begin{enumerate}
\item Let $\beta_{T-1}(i) = 1$, for $i=0,1,\ldots,N-1$.
\item For $t=T-2,T-3,\ldots,0$ and $i=0,1,\ldots,N-1$ compute
$$
  \beta_t(i) = \sum_{j=0}^{N-1} a_{ij}b_j(\O_{t+1})\beta_{t+1}(j) .
$$
\end{enumerate}

Next, for $t=0,1,\ldots,T-1$ and $i=0,1,\ldots,N-1$, define
$$
  \gamma_t(i) = P(x_t=q_i\,|\,\O,\lambda) .
$$
Since~$\alpha_t(i)$ measures the relevant probability 
up to time~$t$ and~$\beta_t(i)$ measures the relevant
probability after time~$t$, we have
$$
  \gamma_t(i) = \frac{\alpha_t(i)\beta_t(i)}{P(\O\,|\,\lambda)} .
$$
As noted above, the denominator~$P(\O\,|\,\lambda)$ 
is obtained by summing~$\alpha_{T-1}(i)$ over~$i$.
From the definition of~$\gamma_t(i)$ 
it follows that the most likely state at time~$t$ is
the state~$q_i$ for which~$\gamma_t(i)$ is 
maximum, where the maximum is taken over
the index~$i$.

\subsubsection{Solution to Problem 3}
We want to adjust the model
parameters to best fit the observations. The sizes of
the matrices ($N$ and~$M$) are specified, but the
elements of~$A$, $B$ and~$\pi$ are to be determined, 
subject to the row stochastic conditions. The fact that
we can efficiently re-estimate the model itself is perhaps the
most impressive aspect of HMMs.

For $t=0,1,\ldots,T-2$ and $i,j\in\{0,1,\ldots,N-1\}$,
define ``di-gammas'' as
$$
  \gamma_t(i,j) = P(x_t=q_i,x_{t+1}=q_j\,|\,\O,\lambda) .
$$
Then~$\gamma_t(i,j)$ is the probability of being in state~$q_i$ 
at time~$t$ and transiting to state~$q_j$ at time~$t+1$.
The di-gammas can be written in terms 
of~$\alpha$, $\beta$, $A$ and~$B$ as
$$
  \gamma_t(i,j) = \frac{\alpha_t(i)a_{ij}b_j(\O_{t+1})\beta_{t+1}(j)}{
                    P(\O\,|\,\lambda)} .
$$
For $t=0,1,\ldots,T-2$, the~$\gamma_t(i)$ and~$\gamma_t(i,j)$
are related by
$$
  \gamma_t(i) = \sum_{j=0}^{N-1}\gamma_t(i,j) .
$$

Given the~$\gamma$ and di-gamma,
the model~$\lambda=(A,B,\pi)$ can be re-estimated as follows.
\begin{enumerate}
\item For $i=0,1,\ldots,N-1$, let
  \begin{equation}\label{eq:reestPi}
    \pi_i = \gamma_0(i)
  \end{equation}
\item For $i=0,1,\ldots,N-1$ and $j=0,1,\ldots,N-1$, compute
  \begin{equation}\label{eq:reestA}
    a_{ij} = \sum_{t=0}^{T-2}\gamma_t(i,j) \bigg/
                    \sum_{t=0}^{T-2}\gamma_t(i) .
  \end{equation}
\item For $j=0,1,\ldots,N-1$ and $k=0,1,\ldots,M-1$, compute
  \begin{equation}\label{eq:reestB}
    b_j(k) = \hspace*{-0.3in}\sum_{\substack{t\in\{0,1,\ldots,T-1\}\\ \O_t=k}}
               \hspace*{-0.3in}\gamma_t(j)\bigg/
               \sum_{t=0}^{T-1}\gamma_t(j) .
  \end{equation}
\end{enumerate}

The numerator of the re-estimated~$a_{ij}$ 
is the expected
number of transitions from state~$q_i$ to state~$q_j$,
while the denominator is the expected number of
transitions from~$q_i$ to any state. The ratio of these two quantities
is the probability of transiting from state~$q_i$
to state~$q_j$, which is our best estimate for~$a_{ij}$. 

The numerator of the re-estimated~$b_j(k)$ is
the expected number of times the model is in state~$q_j$
and we observe symbol~$k$, while the denominator is the
expected number of times the model is in state~$q_j$.
The ratio is the probability of observing~$k$, 
given that the model is in state~$q_j$. This ratio
is our best estimate for~$b_j(k)$.

Re-estimation is an iterative process.
Typically, we initialize~$\lambda=(A,B,\pi)$
with random values
such that~$\pi_i\approx 1/N$ and~$a_{ij}\approx 1/N$ 
and~$b_j(k)\approx 1/M$. It is critical that~$A$,
$B$ and~$\pi$ be randomized, since exactly uniform values will
result in a local maximum and the model will fail to climb.
The matrices~$\pi$, $A$ and~$B$ must be initialized to be
row stochastic.

The solution to Problem~3 can be summarized as follows.
\begin{enumerate}
\item Initialize,~$\lambda=(A,B,\pi)$.
\item Compute $\alpha_t(i)$, $\beta_t(i)$, 
$\gamma_t(i,j)$ and~$\gamma_t(i)$.
\item Re-estimate the model~$\lambda=(A,B,\pi)$.
\item If $P(\O\,|\,\lambda)$ increases, goto 2. 
\end{enumerate}
In Section~\ref{chap:Implementation}, we discuss how we use HMMs in the
experiments considered in this paper.

\section{Profile Hidden Markov Models}\label{sec:phmmOverview}

In this section we introduce Profile Hidden Markov Models.
Note that in a Hidden Markov Model, the positional information within 
the observation sequence is not relevant. A PHMM can be viewed
as an HMM that takes this positional information into account.
In addition, a PHMM accounts for possible insertions and deletions within the
observation sequence, which is not the case with a standard HMM.

Not surprisingly, PHMMs are more complex to describe than HMMs.
Hence, we do not provide quite as much detail in this 
section as we did in the previous section.

\subsection{Overview and Notation\label{sect:PHMMnotation}}

First, we consider state transitions within a PHMM.
That is, we want to determine the PHMM equivalent of the~$A$ matrix in an HMM. 



\begin{figure*}[htb]
  \begin{center}
    \begin{tikzpicture}[scale=0.9]
    
    \draw[thick,color=blue] (0,0) rectangle (1,1);
    \draw[thick,color=blue] (2.5,0) rectangle (3.5,1);
    \draw[thick,color=blue] (5,0) rectangle (6,1);
    \draw[thick,color=blue] (7.5,0) rectangle (8.5,1);
    \draw[thick,color=blue] (10,0) rectangle (11,1);
    \draw[thick,color=blue] (12.5,0) rectangle (13.5,1);
    
    \draw[thick,color=red,rotate around={45:(0.5,3)}] (0,2.5) rectangle (1.0,3.5);
    \draw[thick,color=red,rotate around={45:(3,3)}] (2.5,2.5) rectangle (3.5,3.5);
    \draw[thick,color=red,rotate around={45:(5.5,3)}] (5,2.5) rectangle (6,3.5);
    \draw[thick,color=red,rotate around={45:(8,3)}] (7.5,2.5) rectangle (8.5,3.5);
    \draw[thick,color=red,rotate around={45:(10.5,3)}] (10,2.5) rectangle (11,3.5);

    \draw[thick,color=green] (3,5.5) circle (0.575);
    \draw[thick,color=green] (5.5,5.5) circle (0.575);
    \draw[thick,color=green] (8,5.5) circle (0.575);
    \draw[thick,color=green] (10.5,5.5) circle (0.575);
   
    \draw[thick,color=blue,->] (1,0.5) -- (2.5,0.5);
    \draw[thick,color=blue,->] (3.5,0.5) -- (5,0.5);
    \draw[thick,color=blue,->] (6,0.5) -- (7.5,0.5);
    \draw[thick,color=blue,->] (8.5,0.5) -- (10,0.5);
    \draw[thick,color=blue,->] (11,0.5) -- (12.5,0.5);

    \draw[thick,color=green,->] (3.57,5.5) -- (4.93,5.5);
    \draw[thick,color=green,->] (6.07,5.5) -- (7.43,5.5);
    \draw[thick,color=green,->] (8.57,5.5) -- (9.93,5.5);

    \draw[thick,color=green,->] (3.25,5.0) -- (5.25,1.0);
    \draw[thick,color=green,->] (5.75,5.0) -- (7.75,1.0);
    \draw[thick,color=green,->] (8.25,5.0) -- (10.25,1.0);
    \draw[thick,color=green,->] (10.75,5.0) -- (12.75,1.0);

    \draw[thick,color=green,->] (3.0,4.925) -- (3.0,3.7);
    \draw[thick,color=green,->] (5.5,4.925) -- (5.5,3.7);
    \draw[thick,color=green,->] (8.0,4.925) -- (8.0,3.7);
    \draw[thick,color=green,->] (10.5,4.925) -- (10.5,3.7);

    \draw[thick,color=blue,->] (0.5,1) -- (0.5,2.3);
    \draw[thick,color=blue,->] (3,1) -- (3,2.3);
    \draw[thick,color=blue,->] (5.5,1) -- (5.5,2.3);
    \draw[thick,color=blue,->] (8,1) -- (8,2.3);
    \draw[thick,color=blue,->] (10.5,1) -- (10.5,2.3);

    \draw[thick,color=blue,->] (0.75,1) -- (2.75,5.0);
    \draw[thick,color=blue,->] (3.25,1) -- (5.25,5.0);
    \draw[thick,color=blue,->] (5.75,1) -- (7.75,5.0);
    \draw[thick,color=blue,->] (8.25,1) -- (10.25,5.0);
    
    \draw[thick,color=red,->] (0.85,2.66) -- (2.5,0.75);
    \draw[thick,color=red,->] (3.35,2.66) -- (5,0.75);
    \draw[thick,color=red,->] (5.85,2.66) -- (7.5,0.75);
    \draw[thick,color=red,->] (8.35,2.66) -- (10,0.75);
    \draw[thick,color=red,->] (10.85,2.66) -- (12.5,0.75);

    \draw[thick,color=red,->] (0.85,3.35) -- (2.5,5.25);
    \draw[thick,color=red,->] (3.35,3.35) -- (5,5.25);
    \draw[thick,color=red,->] (5.85,3.35) -- (7.5,5.25);
    \draw[thick,color=red,->] (8.35,3.35) -- (10,5.25);
 
    \draw[thick,color=red,->] (0.15,2.66) .. controls (-0.75,2.2) and (-0.75,3.8) .. (0.15,3.34);
    \draw[thick,color=red,->] (2.65,2.66) .. controls (1.75,2.2) and (1.75,3.8) .. (2.65,3.34);
    \draw[thick,color=red,->] (5.15,2.66) .. controls (4.25,2.2) and (4.25,3.8) .. (5.15,3.34);
    \draw[thick,color=red,->] (7.65,2.66) .. controls (6.75,2.2) and (6.75,3.8) .. (7.65,3.34);
    \draw[thick,color=red,->] (10.15,2.66) .. controls (9.25,2.2) and (9.25,3.8) .. (10.15,3.34);

    \node at (0.5,0.5) {\small begin};
    \node at (3,0.5) {$M_1$};
    \node at (5.5,0.5) {$M_2$};
    \node at (8,0.5) {$M_3$};
    \node at (10.5,0.5) {$M_4$};
    \node at (13,0.5) {\small end};

    \node at (0.5,3) {$I_0$};
    \node at (3,3) {$I_1$};
    \node at (5.5,3) {$I_2$};
    \node at (8,3) {$I_3$};
    \node at (10.5,3) {$I_4$};

    \node at (3,5.5) {$D_1$};
    \node at (5.5,5.5) {$D_2$};
    \node at (8,5.5) {$D_3$};
    \node at (10.5,5.5) {$D_4$};

    \end{tikzpicture}
  \end{center}
  \caption{Profile Hidden Markov Model\label{fig:PHMM}}
\end{figure*}
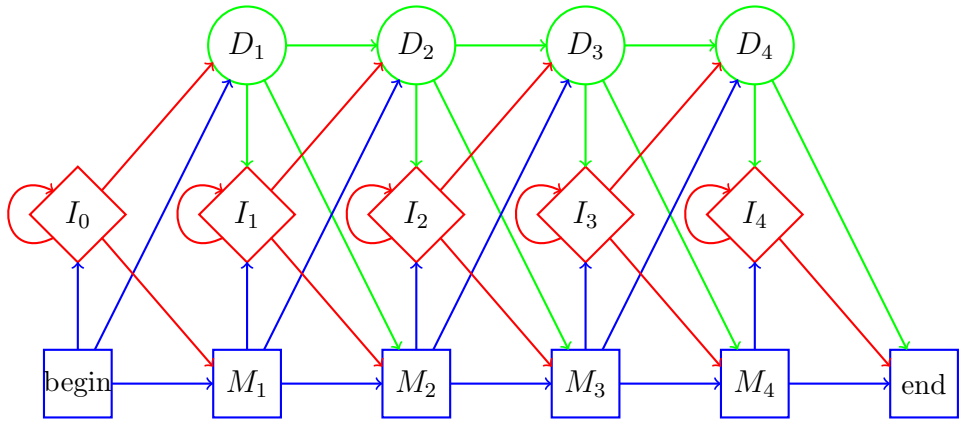

Generically, a PHMM includes match ($M$), insert ($I$), 
and delete ($D$) states, as 
illustrated in Figure~\ref{fig:PHMM}. This illustration is essentially the
PHMM equivalent of the hidden states (and transitions) in the 
HMM illustration in Figure~\ref{fig:hmm}.
That is, the complicated illustration in Figure~\ref{fig:PHMM} is just the 
state transitions of the PHMM, and does not directly include information
related to the emissions (i.e., observations). 
That is, nothing in this figure that
corresponds to the~$B$ matrix in the HMM.

For a PHMM, the standard notation~\cite{DEKM} is 
summarized in Table~\ref{tab:PHMMnotation}.
\begin{table}[htb]
  \begin{center}
  \caption{PHMM Notation\label{tab:PHMMnotation}}
  \begin{tabular}{cl} \hline\hline
    notation & \hspace*{0.5in}explanation\\ \hline
    $X$ & sequence of emitted symbols, $x_1,x_2,\ldots,x_i$\\
    $N$ & number of states\\
    $M$ & match states, $M_1,M_2,\ldots,M_N$\\
    $I$ & insert states, $I_1,I_2,\ldots,I_N$\\
    $D$ & delete states, $D_1,D_2,\ldots,D_N$\\
    $\pi$ & initial state distribution\\
    $A$ & state transition probability matrix\\
    $a_{M_iM_{i+1}}$ & transition probability from~$M_i$ to~$M_{i+1}$\\
    $E$ & emission probability matrix\\
    $\e_{M_i}(k)$ & emission probability of symbol~$k$ at state~$M_i$ \\
    $\lambda$ & the PHMM, $\lambda=(A,E,\pi)$\\ \hline\hline
  \end{tabular}
  \end{center}
\end{table}
It may be instructive to compare the PHMM notation in Table~\ref{tab:PHMMnotation}
to the HMM notation that is given in 
Table~\ref{tab:HMMnotation}.

Next, we turn our attention to generating pairwise alignments.
Then we consider the process of generating a multiple sequence alignment (MSA) from a collection
of pairwise alignments, and we discuss the process used to generate the PHMM
matrices from an MSA. As mentioned above, generating the MSA is the 
most challenging step in the training process. Finally, we consider the scoring 
process using a PHMM.
The scoring procedure for a PHMM uses a similar process as is used in an HMM.

\subsection{Pairwise Alignment\label{sect:PHMMpair}}

We require a method to align a pair of sequences from the training set. 
Ideally, we would like to globally align a pair of sequences, that is, we want an
alignment that accounts for all elements in both sequences. However, 
we also want to minimize the number of gaps that are inserted, since gaps
tend to weaken the resulting PHMM by making it more generic. 
By using a local alignment strategy
instead of a global approach, we can often significantly reduce the number
of gaps. The tradeoff is that a local alignment strategy does
not utilize all of the information available in the sequences.

To simplify the local alignment problem, we only consider such
alignments where the initial and ending parts of the sequences 
can remain unaligned.
For example, suppose that we want to align the sequences
$$
  \mbox{\tt CBCBJILIIJEJE} \mbox{ and } \mbox{\tt GCBJIIIJJEG} .
$$
In Table~\ref{tab:PHMMglobal} we give a global alignment,
along with a local alignment, where the initial and final parts of the sequences
are not aligned. Note that we use ``{\tt -}'' to represent
an inserted gap, while ``{\tt *}'' is an omitted symbol,
that is, a symbol omitted from
consideration in the local alignment. Also, ``{\tt |}'' indicates
the corresponding elements are aligned. 

\begin{table}[htb]
  \begin{center}
  \caption{Global versus Local Alignment\label{tab:PHMMglobal}}
  \begin{tabular}{ll} \hline\hline
  unaligned sequences & {\tt CBCBJILIIJEJE} \\ 
                                     & {\tt GCBJIIIJJEG} \\ \hline
  global alignment & {\tt -CBCBJILIIJEJE-}\\
                             & {\tt \ |\ \ |||\ |||\ ||} \\
                             & {\tt GC--BJI-IIJ-JEG}\\ \hline
  local alignment & {\tt ***CBJILII-JE**}\\
                           & {\tt \ \ \ ||||\ ||\ ||} \\
                           & {\tt ***CBJI-IIJJE**}\\ \hline\hline
  \end{tabular}
  \end{center}
\end{table}

For the global alignment
in Table~\ref{tab:PHMMglobal}, we are able to align~9 out of~15 of the positions 
(60\%), while for the local alignment, we have aligned~8 of the~10 positions 
(80\%) under consideration. Consequently, model based on this local alignment is
likely to be stronger than a model based on the global alignment. Therefore,
we consider local alignments of the type illustrated in Table~\ref{tab:PHMMglobal}.
However, to simplify the presentation, in the remainder of this section, we 
illustrate global alignments. To obtain optimal local alignments, we could simply repeat the 
alignment process several times, omitting symbols at the beginning 
and end of the sequences.

To construct a pairwise alignment, it is standard practice to use dynamic programming.
In the context of sequence alignment,
we specify an~$n\times n$ substitution matrix~$S$, where~$n$ is the number of symbols under 
consideration, and a gap penalty function~$g$. The substitution matrix specifies the
cost associated with aligning various symbols with each other, while~$g$ specifies
the penalty incurred when opening (or extending) a gap.
Once~$S$ and~$g$ have been specified, 
the dynamic program is relatively straightforward. For the sake of brevity,
we omit further details on pairwise alignment.
For more information on sequence alignment within the
context of a PHMM, see Chapter~3 of the tutorial~\cite{Stamp_ML}.

\subsection{Multiple Sequence Alignment\label{sect:PHM_MSA}}

Given a set of training sequences, we construct pairwise alignments for
all pairs. From these pairwise alignments, we want to construct
a multiple sequence alignment. Constructing the MSA is
essentially the training process for a PHMM.

For efficiency, a progressive alignment strategy is generally used to construct the MSA.
That is, we start with one pair of aligned sequence and merge it with another
aligned pair, and merge that result with another, and so on. In this way, at each step
we include more sequences into the MSA, until all training sequences
have been used. The advantage of a progressive approach is that it is far more
efficient than trying to align multiple sequences simultaneously. The
disadvantage of a progressive alignment strategy is that it is likely to be unstable,
in the sense that the order in which the pairwise alignments are considered can
have a major impact on the resulting MSA. Specifically, the order can affect the number
of gaps that are inserted into the MSA, and it is always desirable to minimize gaps.

There are many progressive alignment techniques available.
Here, we discuss an approach based on the Feng-Dolittle algorithm,
as used in~\cite{bib28}. 
To construct an MSA, we proceed as follows, where
we assume that we are given as set of~$n$ training sequences, 
a substitution matrix~$S$ and a gap penalty function~$g$.
\begin{enumerate}
\item Compute pairwise alignments for all~$n \choose 2$ pairs of sequences in the
training set, using dynamic programming.
For each pairwise alignment, we obtain a score from the dynamic program.
\item Select a set of~$n-1$ pairwise alignments that 
includes all~$n$ sequences from the original training set, and maximizes
the sum of the pairwise alignment scores.
\item Generate a minimum spanning tree (using Prim's algorithm) for these~$n-1$
pairwise alignments, based on the pairwise alignment scores.
\item Add pairwise alignments to the MSA based on the
spanning tree (from highest score to lowest score), inserting
additional gaps as needed. The gap penalty function~$g$ is used when inserting gaps.
\end{enumerate}

For example, suppose we have~10 training sequences which result in the 
pairwise alignment scores in Table~\ref{tab:PHMMalignScores},
where element~$(i,j)$ is the score obtained for the pairwise alignment
of sequence~$i$ with sequence~$j$.

\begin{table}[htb]
\begin{center}
  \caption{Example Pairwise Alignment Scores\label{tab:PHMMalignScores}}
  \begin{tabular}{r|rrrrrrrrrr}\hline\hline
     & 1 & 2 & 3 & 4 & 5 & 6 & 7 &8 & 9 & 10\\ \hline
  1 & --- & 85 & 63 & 74 & 70 & 84 & 61 & 57 & 62 & 70\\
  2 & 85 & --- & 79 & 73 & 66 & 59 & 94 & 61 & 59 & 51\\
  3 & 63 & 79 & --- & 75 & 68 & 60 & 55 & 85 & 52 & 65\\
  4 & 74 & 73 & 75 & --- & 105 & 54 & 60 & 78 & 59 & 53\\
  5 & 70 & 66 & 68 & 105 & --- & 40 & 61 & 79 & 58 & 39\\
  6 & 84 & 59 & 60 & 54 & 40 & --- & 68 & 45 & 75 & 78\\
  7 & 61 & 94 & 55 & 60 & 61 & 68 & --- & 64 & 72 & 42\\
  8 & 57 & 61 & 85 & 78 & 79 & 45 & 64 & --- & 50 & 70\\
  9 & 62 & 59 & 52 & 59 & 58 & 75 & 72 & 50 & --- & 81\\
10 & 70 & 51 & 65 & 53 & 39 &78 & 42 & 70 & 81 & ---\\ \hline\hline
\end{tabular}
\end{center}
\end{table}

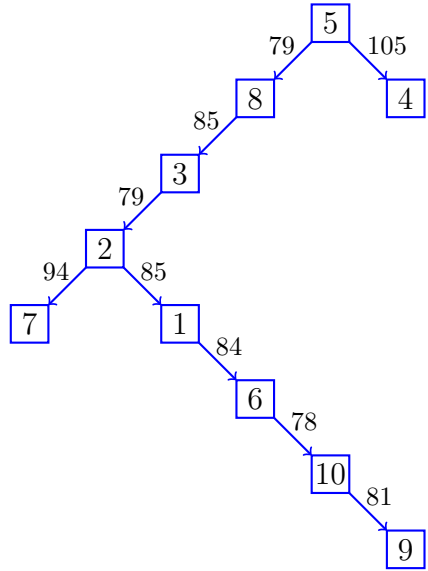
\begin{figure}[htb]
  \begin{center}
  \def\ff{\footnotesize}
    \begin{tikzpicture}[scale=1.0]
    
    \draw[thick,color=blue] (4,7) rectangle (4.5,7.5);	\draw[thick,color=blue] (5,6) rectangle (5.5,6.5);
    \draw[thick,color=blue] (3,6) rectangle (3.5,6.5);
    \draw[thick,color=blue] (2,5) rectangle (2.5,5.5);
    \draw[thick,color=blue] (1,4) rectangle (1.5,4.5);
    \draw[thick,color=blue] (0,3) rectangle (0.5,3.5);	\draw[thick,color=blue] (2,3) rectangle (2.5,3.5);
   										\draw[thick,color=blue] (3,2) rectangle (3.5,2.5);
    										\draw[thick,color=blue] (4,1) rectangle (4.5,1.5);
    										\draw[thick,color=blue] (5,0) rectangle (5.5,0.5);

    \draw[thick,color=blue,->] (4,7) -- (3.5,6.5);	\draw[thick,color=blue,->] (4.5,7) -- (5,6.5);
    \draw[thick,color=blue,->] (3,6) -- (2.5,5.5);
    \draw[thick,color=blue,->] (2,5) -- (1.5,4.5);
    \draw[thick,color=blue,->] (1,4) -- (0.5,3.5);	\draw[thick,color=blue,->] (1.5,4) -- (2,3.5);
									\draw[thick,color=blue,->] (2.5,3) -- (3,2.5);
									\draw[thick,color=blue,->] (3.5,2) -- (4,1.5);
									\draw[thick,color=blue,->] (4.5,1) -- (5,0.5);

    \node at (4.25,7.25) {$5$};	\node at (5.25,6.25) {$4$};
    \node at (3.25,6.25) {$8$};
    \node at (2.25,5.25) {$3$};
    \node at (1.25,4.25) {$2$};
    \node at (0.25,3.25) {$7$}; 	\node at (2.25,3.25) {$1$};
    						\node at (3.25,2.25) {$6$};
    						\node at (4.25,1.25) {$10$};
    						\node at (5.25,0.25) {$9$};

    \node at (3.6,6.95) {\ff $79$};	\node at (5,6.95) {\ff $105$};
    \node at (2.6,5.95) {\ff $85$};
    \node at (1.6,4.95) {\ff $79$};
    \node at (0.6,3.95) {\ff $94$};	\node at (1.9,3.95) {\ff $85$};
    							\node at (2.9,2.95) {\ff $84$};
    							\node at (3.9,1.95) {\ff $78$};
    							\node at (4.9,0.95) {\ff $81$};

    \end{tikzpicture}
  \end{center}
  \caption{Minimum Spanning Tree\label{fig:PHMM_MST}}
\end{figure}

We then select a set of~$9$ of these pairs so that each sequence~$1,2,\dots,10$ 
is included at least once, while maximizing the sum of the scores. 
Using the scores in Table~\ref{tab:PHMMalignScores},
the set 
$$
  \{(4,5), (2,7), (1,2), (3,8), (1,6), (9,10), (2,3), (5,8), (6,10)\} 
$$
satisfies these requirements. The minimum spanning tree 
corresponding to these pairwise alignments
is given in Figure~\ref{fig:PHMM_MST},
where the nodes are labeled with the training sequence number, and the edges
are labeled with pairwise alignment scores.

Using this procedure, we construct a minimum spanning tree
that includes every sequence in the training set. This spanning tree will be used to 
generate the MSA which, in turn, is used to generate the PHMM.

To generate an MSA from a minimum spanning tree, such as that in Figure~\ref{fig:PHMM_MST},
we start with the pairwise alignment at the root. Then we traverse the tree, adding the specified new
sequence into the alignment at each step. For the example in Figure~\ref{fig:PHMM_MST}, we consider
the pairwise alignments in the order
$$
  (5,4),(5,8),(8,3),(3,2),(2,7),(2,1),(1,6),(6,10),\mbox{ and } (10,9) .
$$ 
With the exception of the initial pair, one new sequence is added to
the MSA at each step.

Again referring to the example in Figure~\ref{fig:PHMM_MST},
one step in the MSA construction is given in Table~\ref{tab:PHMMsnapshot}, 
At this particular step, we have already included sequences 5, 4, 8, and 3, based
on the pairwise alignments
$$
  (5,4),(5,8),\mbox{ and }(8,3) 
$$ 
and we are adding sequence~2 based on the pairwise alignment~$(3,2)$.
Note that at these intermediate steps, we have used ``{\tt +}'' to represent gaps that
are inserted to align the new sequence with the existing MSA, as opposed to ``{\tt -}'' which
we use for gaps in the pairwise alignments. The final MSA appears in Table~\ref{tab:PHMMfinal}, 
where all gaps are represented by ``{\tt -}'' symbols.

\begin{table}[htb]
\begin{center}
\def\h{\hspace*{0.25in}}
\caption{Snapshots of MSA Construction\label{tab:PHMMsnapshot}}
{\small
\begin{tabular}{c|l}\hline\hline
\multicolumn{2}{c}{MSA at intermediate step}\\ \hline
seq. & \multicolumn{1}{c}{alignment} \\ \hline
5 & \verb~CDABBAFCDB1AAEAA+CEDA+EQ+CDABABABALF4~\\ 
4 & \verb~2AABBAFCDABA+EAABCEDCDEQFCDABA+APALF4~\\ 
8 & \verb~++AABA+CDB+AAEAA+CEDCDEQ+CDABPBA+ABF4~\\ 
3 &  \verb~A+ABBAFCDABA+EAA+CEDCDEQA++ABFBAN++F4~\\ 
\hline 
\multicolumn{2}{c}{ }\\ 
\multicolumn{2}{c}{Next pairwise alignment}\\ \hline
seq. & \multicolumn{1}{c}{alignment} \\ \hline
2 &  \verb~A-ABNBAFCD-BAAEAABCEDA-EQ-CDABAB--BAF4~\\ 
3 &  \verb~A+AB-BAFCDABA+EAA+CEDCDEQA++ABFBAN++F4~\\ 
\hline 
\multicolumn{2}{c}{ }\\ 
\multicolumn{2}{c}{MSA after including sequence 2}\\ \hline
seq. & \multicolumn{1}{c}{alignment} \\ \hline
5 &  \verb~CDAB+BAFCDB1AAEAA+CEDA+EQ+CDABABABALF4~\\ 
4 &  \verb~2AAB+BAFCDABA+EAABCEDCDEQFCDABA+APALF4~\\ 
8 &  \verb~++AA+BA+CDB+AAEAA+CEDCDEQ+CDABPBA+ABF4~\\ 
3 &  \verb~A+AB+BAFCDABA+EAA+CEDCDEQA++ABFBAN++F4~\\ 
2 &  \verb~A+ABNBAFCD+BAAEAABCEDA+EQ+CDABAB++BAF4~\\ 
\hline\hline
\end{tabular}
}
\end{center}
\end{table}

\begin{table}[htb]
\begin{center}
\def\h{\hspace*{0.25in}}
\caption{Final MSA\label{tab:PHMMfinal}}
\vglue -10pt
{\small
\begin{tabular}{c|l}\hline\hline
seq. & \multicolumn{1}{c}{alignment} \\ \hline
 \z1 &  \verb~A-AB-BAFCD-B-AAEA0ACEDA-EQ---A-ABCDBALF4~\\ 
 \z2 &  \verb~A-ABNBAFCD-B-AAEAABCEDA-EQ-CDABAB--BA-F4~\\ 
 \z3 &  \verb~A-AB-BAFCDAB-A-EAA-CEDCDEQA--ABFBAN---F4~\\ 
 \z4 &  \verb~2AAB-BAFCDAB-A-EAABCEDCDEQFCDABA-APAL-F4~\\ 
 \z5 &  \verb~CDAB-BAFCDB1-AAEAA-CEDA-EQ-CDABABABAL-F4~\\ 
 \z6 &  \verb~CDABAAA----B-A-EA-ACEDCDEQ---A-ABCD-A-F4~\\ 
 \z7 &  \verb~CDAB--A-CDAB-A-EAA-CEDA-EQ-CDABCDCDAA-F4~\\ 
 \z8 &  \verb~--AA-BA-CDB--AAEAA-CEDCDEQ-CDABPBA-AB-F4~\\ 
 \z9 &  \verb~CDAB--RBAFABPAAEA-ACEDCDEQAABCDAFAL---F4~\\ 
10 &  \verb~A-ABAA-----B-AAEA-ACEDCDEQAABAFA------F4~\\ 
\end{tabular}
}
\end{center}
\end{table}

\subsection{PHMM from MSA}

A PHMM is determined directly from an MSA, that is, the 
probabilities in the matrices in~$\lambda=(A,E,\pi)$ are determined
from the MSA. The process might look somewhat complex, but
it is actually straightforward, since probabilities are based only
on counts of elements that appear in the MSA.
First, we consider the emission probability matrix~$E$, then
we turn our attention to the state transition matrix~$A$.

Recall that the PHMM includes match ($M$), insert ($I$), 
and delete ($D$) states.
We define ``conservative'' columns of the MSA as those
for which half or less of the elements are gaps. Conservative columns
correspond to match states of the PHMM. In contrast, if the majority of
elements in a column are gaps, the column represents an insert state.
The delete states will be considered later.

Consider the simple MSA in Table~\ref{tab:PHMM_MSAex},
which only includes the four symbols, C, E, G, and J. 
Note that columns~1, 2 and~6 are conservative, and hence
correspond to match states. Columns~3, 4, and~5 are not conservative.
Consecutive non-conservative columns are
treated as a single insert state. 

\begin{table}[htb]
  \begin{center}
  \caption{MSA Example\label{tab:PHMM_MSAex}}
     \setlength{\unitlength}{3.66pt}
    \begin{picture}(100,34)(4,-2)
      \thicklines
      
      \put(23,32.85){\line(1,0){29.5}}
      \put(23,32.25){\line(1,0){29.5}}
      
      \put(23,27.5){\line(1,0){29.5}}
      \put(23,2.5){\line(1,0){29.5}}

      \put(23,-2.0){\line(1,0){29.5}}
      \put(23,-2.6){\line(1,0){29.5}}

      \put(24,22){E}
      \put(24,18){E}
      \put(24.25,14){--}
      \put(24,10){E}
      \put(24,6){E}
      \put(24.5,-0.75){1}
      \put(23.25,29){$M_1$}
      
      \put(29,22){C}
      \put(29,18){C}
      \put(29,14){C}
      \put(29,10){G}
      \put(29,6){G}
      \put(29.5,-0.75){2}
      \put(28.5,29){$M_2$}
      
      \put(33,4.5){\dashbox(14.5,21){}}
      \put(39.25,29){$I_2$}

      \put(34.25,22){--}
      \put(34.25,18){--}
      \put(34,14){G}
      \put(34.25,10){--}
      \put(34.25,6){--}
      \put(34.5,-0.75){3}
      
      \put(39.25,22){--}
      \put(39,18){E}
      \put(39,14){E}
      \put(39.25,10){--}
      \put(39.25,6){--}
      \put(39.5,-0.75){4}
      
      \put(44.25,22){--}
      \put(44.25,18){--}
      \put(44,14){J}
      \put(44,10){J}
      \put(44.25,6){--}
      \put(44.5,-0.75){5}
      
      \put(49.25,22){--}
      \put(49,18){G}
      \put(49,14){G}
      \put(49,10){G}
      \put(49,6){G}
      \put(49.5,-0.75){6}
      \put(48,29){$M_3$}

 
     \end{picture}

  \end{center}
\end{table}

Emissions occur at match and insert states. The probabilities in the~$E$ matrix
are determined from the MSA based on the counts in each state. 
Referring to the MSA in Table~\ref{tab:PHMM_MSAex}, 
for column~1 we have
\begin{equation}\label{eq:PHMMno-add-one}
	\e_{M_1}(\E) = 4/4,\ \e_{M_1}(\G) = e_{M_1}(\C) = e_{M_1}(\J) = 0/4
\end{equation}
since all~4 of the (non-gap) symbols that appear are~E.

Any model that includes probabilities of zero is prone to overfit the
training data, since ``nearby'' sequences are eliminated. 
Several standard methods are available to eliminate zero probabilities.
Here, we employ the  ``add-one rule''~\cite{DEKM} which, consists of
adding one to each numerator and, so as to maintain probabilities,
also adding the total number of symbols to each denominator. Since there are four distinct 
symbols in our example, applying the add-one rule to the probabilities in 
equation~\eref{eq:PHMMno-add-one} yields 
$$
	\e_{M_1}(\E) = 5/8,\ \e_{M_1}(\G) = \e_{M_1}(\C) = \e_{M_1}(\J) = 1/8 .
$$

For the insert state~$I_2$, the natural probabilities are
$$
	 \e_{I_2}(\E) = 2/5,\ \e_{I_2}(\G) = 1/5,\ \e_{I_2}(\C) = 0/5,\ \e_{I_2}(\J) = 2/5
$$
which come from the ratios of the emitted symbols in the dashed box
in Table~\ref{tab:PHMM_MSAex}. Using the add-one rule, the
insert state probabilities become
$$
	 \e_{I_2}(\E) = 3/9,\ \e_{I_2}(\G) = 2/9,\ \e_{I_2}(\C) = 1/9,\ \e_{I_2}(\J) = 3/9 .
$$
The emission probabilities for the example in Table~\ref{tab:PHMM_MSAex}, 
with the add-one rule applied, are given Table~\ref{tab:PHMMemprob}.
Note that for states for which we have no information (e.g., $I_1$ in
this example), we assume a uniform distribution.

\begin{table}[htb]
\caption{Emission Probabilities for the MSA in 
Table~\ref{tab:PHMM_MSAex}\label{tab:PHMMemprob}}
\begin{center}\def\hhh{\hspace*{0.5in}}
\begin{tabular}{rr} \hline\hline
$\e_{M_1}(\E) = 5/8$ & \hhh $\e_{I_1}(\E) = 1/4$ \\
$\e_{M_1}(\G) = 1/8$ & \hhh $\e_{I_1}(\G) = 1/4$ \\
$\e_{M_1}(\C) = 1/8$ & \hhh $\e_{I_1}(\C) = 1/4$ \\
$\e_{M_1}(\J) = 1/8$ & \hhh $\e_{I_1}(\J) = 1/4$ \\ 
\hline
$\e_{M_2}(\E) = 1/9$ & \hhh $\e_{I_2}(\E) = 3/9$ \\
$\e_{M_2}(\G) = 3/9$ & \hhh $\e_{I_2}(\G) = 2/9$ \\
$\e_{M_2}(\C) = 4/9$ & \hhh $\e_{I_2}(\C) = 1/9$ \\
$\e_{M_2}(\J) = 1/9$ & \hhh $\e_{I_2}(\J) = 3/9$ \\ 
\hline
$\e_{M_3}(\E) = 1/8$ & \hhh $\e_{I_3}(\E) = 1/4$ \\
$\e_{M_3}(\G) = 5/8$ & \hhh $\e_{I_3}(\G) = 1/4$ \\
$\e_{M_3}(\C) = 1/8$ & \hhh $\e_{I_3}(\C) = 1/4$ \\
$\e_{M_3}(\J) = 1/8$ & \hhh $\e_{I_3}(\J) = 1/4$ \\ 
\hline\hline
\end{tabular}
\end{center}
\end{table}

Intuitively, it might seem that the more sequences we use for training,
the stronger the resulting model, since more sequences provide more information. 
However, by using more training sequences,
we are likely to have more gaps in the resulting MSA, 
which tends to weaken the model.
Consequently, the number of sequences used to generate
the MSA is a critical parameter that we can analyze via experimentation.

Next, we consider the transition probabilities, that is, we show how to derive the
state transition probabilities from an MSA. Again, we illustrate the process
using simple MSA in Table~\ref{tab:PHMM_MSAex}.

Intuitively, the probabilities should be given by
\begin{equation}\label{eq:PHMMtrans}
  a_{mn}  = \frac{\mbox{transitions from state $m$ to state $n$}}{
                           \mbox{total transitions from state $m$ to any state}} .
\end{equation}
As with emission probabilities, we want to avoid probabilities of zero,
so we will use the add-one rule.

Let~$B$ be the begin state. 
Then, ignoring the add-one rule,
\begin{equation}\label{eq:PHMMbm}
  a_{BM_1} = 4/5 
\end{equation}
since four of the five elements in column~1 are matches,
which implies that the probability of transitioning from the begin
state to~$M_1$ is~$4/5$.. 
Similarly,
\begin{equation}\label{eq:PHMMbd}
  a_{BD_1} = 1/5\mbox{\ \ and\ \ } a_{BI_0} = 0/5
\end{equation}
since one element in column~1 represents delete state~$D_1$ 
and insert state~$I_0$ is empty.

As with the emission probability calculations, we 
use the add-one rule. However, instead of adding one for each symbol, 
we add one for each possible transition, namely, match, insert, and delete. 
Thus, using the add-one rule, equations~\eref{eq:PHMMbm} and~\eref{eq:PHMMbd} 
become
$$
  a_{BM_1} = 5/8,\ \ a_{BD_1} = 2/8,\mbox{\ \ and\ \ } a_{BI_0} = 1/8 
$$
respectively.

As with the emission probabilities,
in cases where there is no data, we set the transition probabilities 
to uniform. For example, we have no transitions from~$I_1$, 
and consequently
$$
  a_{I_1M_2} = a_{I_1I_1} =  a_{I_1D_2} = 1/3 .
$$

Next, consider the delete state~$D_1$, 
which corresponds to the ``dash''  in column~1 of Table~\ref{tab:PHMM_MSAex}. 
We see that the transition from~$D_1$ is to a 
match state in column~2. Consequently, 
$$
  a_{D_1M_2} = 2/4,\ \ a_{D_1I_1} = 1/4,\mbox{\ \ and\ \ } 
  a_{D_1D_2} = 1/4 
$$
where we have used the add-one rule.

Now consider~$M_2$. In the bottom row, no letter appears in the boxed
region of the MSA in Table~\ref{tab:PHMM_MSAex} and, 
consequently, for this row, we transition from~$M_2$ to~$M_3$.
Similarly, in the top row, we transition from~$M_2$ to~$D_3$. However, the three middle
rows all go from~$M_2$ to~$I_3$. Therefore, using the add-one rule, we have
$$
  a_{M_2M_3} = 2/8,\ \ a_{M_2D_3} = 2/8,\mbox{\ \ and\ \ } 
  a_{M_2I_3} = 4/8 . 
$$

Finally, we calculate transition probabilities for~$I_2$. Note that there are five symbols
in~$I_2$, and of these, three transition to~$M_3$, namely, the~E in the second row,
the~J in the third row, and the~J in the fourth row. Both of the remaining symbols (G and~E
in the third row) transition to symbols in~$I_2$. Therefore,
$$
  a_{I_2M_3} = 4/8,\ \ a_{I_2I_2} = 3/8,\mbox{\ \ and\ \ } 
  a_{I_2D_3} = 1/8 . 
$$

The complete set of transition probabilities for the MSA in Table~\ref{tab:PHMM_MSAex} 
appears in Table~\ref{tab:PHMMtransprob}. Note that the add-one rule has been applied
in all cases.

\begin{table}[htb]
\caption{Transition Probabilities for the MSA in 
Table~\ref{tab:PHMM_MSAex}\label{tab:PHMMtransprob}}
\begin{center}\def\hh{\hspace*{0.33in}}
\begin{tabular}{lll}\hline\hline
$a_{BM_1}  = 5/8$ & \hh $a_{I_0M_1} = 1/3$ \\
$a_{BI_0}   = 1/8$ & \hh $a_{I_0I_0}  = 1/3$ \\
$a_{BD_1}  = 2/8$ & \hh $a_{I_0D_1} = 1/3$ \\ \hline
$a_{M_1M_2} = 5/7$ & \hh $a_{I_1M_2} = 1/3$ & \hh $a_{D_1M_2} = 2/4$ \\
$a_{M_1I_1}  = 1/7$ & \hh $a_{I_1I_1}  = 1/3$ & \hh $a_{D_1I_1}  = 1/4$ \\ 
$a_{M_1D_2} = 1/7$ & \hh $a_{I_1D_2} = 1/3$ & \hh $a_{D_1D_2} = 1/4$ \\ \hline
$a_{M_2M_3} = 2/8$ & \hh $a_{I_2M_3} = 4/8$ & \hh $a_{D_2M_3} = 1/3$ \\
$a_{M_2I_2}  = 4/8$ & \hh $a_{I_2I_2}  = 3/8$ & \hh $a_{D_2I_2}  = 1/3$ \\
$a_{M_2D_3} = 2/8$ & \hh $a_{I_2D_3} = 1/8$ & \hh $a_{D_2D_3} = 1/3$ \\ \hline
$a_{M_3E}  = 5/6$ & \hh $a_{I_3E}  = 1/2$ & \hh $a_{D_3E}  = 2/3$ \\
$a_{M_3I_3} = 1/6$ & \hh $a_{I_3I_3}  = 1/2$ & \hh $a_{D_3I_3}  = 1/3$ \\ \hline\hline
\end{tabular}
\end{center}
\end{table}

To conclude this section, we summarize the process used
to training a PHMM. Here, we assume that we are given a set of training sequences
and that a substitution matrix~$S$ and gap penalty function~$g$
have been specified.
\begin{enumerate}
\item Construct pairwise alignments for the training sequences using~$S$ and~$g$.
Typically, dynamic programming is used in this step.
\item From the pairwise alignments, construct an MSA. There are many ways to construct an MSA
from pairwise alignments---in this paper, we used a spanning tree and a 
progressive alignment strategy. 
\item Use the MSA to determine the PHMM, that is, the probabilities
that constitute~$\lambda=(A,E,\pi)$ are determined directly from the MSA. To
avoid zero probabilities, use some form of pseudo-counts,
such as the add-one rule.
\end{enumerate}

To score a given sequence against a specific PHMM, we use the forward algorithm.
The PHMM version of the forward algorithm is 
fairly similar to the forward algorithm as used for scoring in HMMs,
so we omit the details here; see~\cite{Stamp_ML} for additional information.

\section{Related Work\label{sect:relatedWork}}

Wong and Stamp~\cite{wongStamp} showed the effectiveness of HMMs
for detecting metamorphic malware based on opcode sequences.
Austin et al.~\cite{duelingHmm} extend this idea to a dueling HMM strategy,
that is, a multi-sensor approach that handles more complex viruses
by using models of both benign and malicious files.
Kalbhor et al.~\cite{kalbhor} highlight how the overhead of the dueling HMM strategy
can be reduced to levels approaching that of Wong and Stamp's approach
by using a tiered analysis.
The central idea in~\cite{kalbhor} is to quickly analyze files with the simpler approach first,
and only use the more expensive multi-sensor approach in the more difficult cases.
Annachhatre et al.~\cite{annachhatre} show how this multi-sensor approach 
can be useful in the classification of malware, based on clustering.

Filiol and Josse~\cite{filiolJosse} discuss statistical testing simulability
and describe how an attacker might use information about the defender's detection strategy
to evade detection.
Madenur Sridhara and Stamp~\cite{mwor} use a
similar strategy in the design of an experimental metamorphic worm MWOR.
The MWOR worm relies primarily on dead code insertion,
which seems to be one of the more effective metamorphic techniques at evading HMM-based detection
that is based on static analysis.

Attaluri et al.~\cite{bib28} use profile hidden Markov models to detect metamorphic malware 
based on static opcode sequences.
The approach works well against certain kinds of metamorphic malware,
but seems to be less effective when the blocks of code are shifted farther apart.
This paper continues this work by instead considering dynamic birthmarks.
While PHMMs trained on static data have not proven particularly useful for malware detection~\cite{bib28},
their positional information seems more likely to be beneficial when training on dynamic birthmarks.

\section{Implementation}\label{chap:Implementation}

In this section, we review the design of each of our systems.
First, we discuss our HMM-based approach, then we turn our 
attention to our PHMM-based technique.

\subsection{Static Analysis Using HMMs}\label{sec:statHmmImpl}

We begin with our system using HMMs and static birthmarks for malware analysis.
As mentioned previously,
this strategy has been used with some success in the literature~\cite{wongStamp,duelingHmm},
but it can be defeated by malware
that inserts dead-code as an obfuscation technique~\cite{mwor}.

Following past research for static HMM analysis~\cite{wongStamp},
we use opcodes as our static birthmarks.
We use IDA Pro to disassemble the malware and generate asm files.
From those files, we extract the mnemonic opcodes for training and testing our models.
Labels, operands, and other details are discarded.

We construct the HMMs from these sequences.
Following the approach taken by Wong and Stamp~\cite{wongStamp},
we use~$N=2$ two hidden states. We also use~800 iterations of the Baum-Welch
re-estimation algorithm when training all of our models.
We use five-fold cross-validation. That is, for each experiment,
the family dataset is partitioned into five
equal subsets, say, $S_1$, $S_2$, $S_3$, $S_4$, and~$S_5$. Then 
we train a model using~$S_1$, $S_2$, $S_3$, and $S_4$, and we
use the resulting model to score~$S_5$ as well as the representative benign set.
This process is repeated five times, with a different subset reserved for testing in each ``fold''.
The use of cross validation serves to smooth any biases in the data,
while also maximizing the number of scores from a given dataset.

\subsection{Dynamic Analysis Using HMMs}

To compare the effectiveness of dynamic versus static analysis,
we also consider a HMM analysis based on dynamic birthmarks.
For the dynamic birthmark, we extract API calls at runtime. 
One useful advantage of this approach
is that API calls would seem to more closely capture the actual functionality of the program than the opcodes,
potentially making it a more resilient tool against various code obfuscation techniques.

To collect the API calls, we use the Buster Sandbox Analyzer~\cite{buster} (BSA).
BSA logs information about file system changes, windows registry changes, port changes, etc.
BSA also executes files in a sandbox so that malware can be run safely.

Once the API call sequences are produced,
they are used for training and testing in the same manner as discussed in 
Section~\ref{sec:statHmmImpl}.
We note that there is an additional parameter in this case, namely,
the amount of time that we run the software to generate the sequences.

\subsection{Dynamic Analysis Using PHMMs}\label{sec:phmmImpl}

Here, we use the same API call sequences as in the dynamic HMM case.
Once we have the sequences of API calls for a given malware family,
we align the sequences.
Some amount of preprocessing is required in this step.
We followed the general approach in~\cite{bib28}, that is,
the API calls are first sorted by the frequency of occurrence.
In our data, the top~36 API calls constitute at least 99.8\%\ of the total API calls for each family
tested. Consequently, we only consider the top~36 API calls, 
with all remaining API calls mapped to a single ``other'' state.
This preprocessing step strengthens the resulting models.

The next step involves creating a substitution matrix
and a gap penalty function.
As described in Section~\ref{sec:phmmOverview},
we use a the Feng-Doolittle algorithm~\cite{feng} 
to create our multiple sequence alignment (MSA).
After constructing the MSA, we use it to build the PHMM, 
as discussed in Section~\ref{sec:phmmOverview}.
Once we have built the PHMM,
we use the forward algorithm to score sequences 
against the PHMM.

\section{Experiments and  Results\label{chap:Experiments and Results}}

In this section, we evaluate the effectiveness of our dynamic and static analysis techniques
and present our results. Specifically, we use a static HMM analysis based on extracted opcodes
as a baseline to compare against both HMMs and PHMMs trained on
dynamically extracted API call sequences.

\subsection{Setup}

For all experiments, we have used the Oracle VM Virtual Box.
The host machine has an Intel(R) Core(TM) i5-3317U CPU@1.7Ghz processor, 6GB RAM,
64-bit system, and Windows 8.1 operating system.
The guest machine used in our experiments was an Oracle Virtual Box 4.3.16 VM
with a base memory of 3310MB, 6 GB RAM, 32 bit system, running Windows 7.
The training and testing of the HMMs and PHMM are done in the host machine
whereas the API calls and opcode extraction are done in the guest machine.

\subsection{Dataset}\label{sec:dataset}

For our malware set, we have considered the following seven different malware families~\cite{NRC}.
\renewcommand{\myitem}[1]{\item {\em #1}}
\begin{itemize}
 \myitem{Cridex}
is a worm that multiplies and spreads through removable drives.
It downloads malicious programs onto the system it has attacked~\cite{bib20}.
\myitem{Harebot}
is a backdoor that affects Windows systems.
The Harebot backdoor enables hackers to gain access to 
the compromised system and steal information~\cite{bib25}.
\myitem{Security Shield}
is fake anti-virus software that falsely claims to protect the system from malware.
Security Shield then tries to convince the user to pay money to 
remove these nonexistent threats~\cite{bib23}.
\myitem{SmartHdd}
targets Windows users.
It tricks the users into thinking that it is a legitimate
hardware monitoring tool (SMART).
It generates fake messages indicating that the hard drive is failing
and tries to convince the user to pay to fix the supposed issues.
It also disables antivirus software in the compromised system~\cite{bib22}.
\myitem{Winwebsec}
is a Windows Trojan that impersonates anti-malware software and displays fake messages stating 
that the user's system has been infected.
Once installed, it tries to convince the user to pay for a fake anti-malware product~\cite{bib21}.
\myitem{Zbot}
is another Trojan. Zbot steals confidential information
such as online credentials~\cite{bib19}.
\myitem{Zeroaccess}
is a Trojan that attacks Windows systems.
It uses a botnet to download other malicious programs onto the compromised system~\cite{bib24}.
\end{itemize}
Table~\ref{tab:malware} lists the number of files from each of the malware families
used in our experiments.

\begin{table}[htb]
{
\caption{Malware Files\label{tab:malware}}
\begin{center}
\begin{tabular}{c|c}\hline\hline
Malware Family & \multicolumn{1}{c}{Number of Files} \\ \hline
Cridex & \z50 \\
Harebot & \z50  \\
Security Shield & \z50  \\
SmartHdd & \z50  \\
Winwebsec & 100  \\
Zbot & 100  \\
Zeroaccess & 100  \\
\hline\hline
\end{tabular}
\end{center}
}
\end{table}

For our representative benign dataset,
we use the~20 Windows executable files listed in
Table~\ref{tab:19}. These executables are
available in Windows 7 or as freeware.

\begin{table}[htbp]
{
\caption{Benign files\label{tab:19}}
\begin{center}
\begin{tabular}{c|lc}\hline\hline
No. & \multicolumn{1}{c}{Benign file} \\ \hline
1 & bdaycontrol.exe  \\
2 & IRATrack.exe \\
3 & MineSweeper.exe  \\
4 & SpiderSolitaire.exe  \\
5 & PurblePlace.exe  \\
6 & FreeCell.exe  \\
7 & Chess.exe  \\
8 & 7zFM.exe  \\
9 & hh.exe  \\
10 & Mahjong.exe  \\
11 & DVDMaker.exe  \\
12 & datewiz.exe  \\
13 & winhlp32.exe  \\
14 & bdaycheck.exe  \\
15 & setup.exe  \\
16 & Countdown Pro.exe \\
17 & unins000.exe  \\
18 & bckgzm.exe  \\
19 & nanoclock.exe  \\
20 & shvlzm.exe  \\ \hline\hline
\end{tabular}
\end{center}
}
\end{table}

\begin{figure*}[htp]
\begin{center}
\begin{tabular}{cc}
\begin{tikzpicture}[scale=0.55]
\begin{axis}[width=0.75\textwidth,height=0.675\textwidth,xmin=-0.0,xmax=51.0,
                   ymin=-30.0,ymax=0.0,
                   legend pos=south east,
                   xlabel={},ylabel={Score}] 
\pgfplotstableread{securityShieldStatic.txt}\mydata;
\addplot[thick,color=blue,mark=square,only marks] 
         table
         [
           x expr=\thisrowno{0}, 
           y expr=\thisrowno{1}
         ] {\mydata};
\pgfplotstableread{securityShieldStatic.txt}\mydata;
\addplot[color=red,mark=diamond*,only marks] 
         table
         [
           x expr=\thisrowno{0}, 
           y expr=\thisrowno{2}
         ] {\mydata};
\legend{{\large Malware}, {\large Benign}}
\end{axis}
\end{tikzpicture}
&
\begin{tikzpicture}[scale=0.55]
\begin{axis}[width=0.75\textwidth,height=0.675\textwidth,xmin=-0.0,xmax=51.0,
                   ymin=-30.0,ymax=0.0,
                   legend pos=south east,
                   xlabel={},ylabel={Score}] 
\pgfplotstableread{securityShieldDynamic.txt}\mydata;
\addplot[thick,color=blue,mark=square,only marks] 
         table
         [
           x expr=\thisrowno{0}, 
           y expr=\thisrowno{1}
         ] {\mydata};
\pgfplotstableread{securityShieldDynamic.txt}\mydata;
\addplot[color=red,mark=diamond*,only marks] 
         table
         [
           x expr=\thisrowno{0}, 
           y expr=\thisrowno{2}
         ] {\mydata};
\legend{{\large Malware}, {\large Benign}}
\end{axis}
\end{tikzpicture}
\\
(a)  Scatterplot for Static Birthmarks & (b) Scatterplot for Dynamic Birthmarks
\\[2ex]
    \begin{tikzpicture}[scale=0.55]
    \begin{axis}[width=0.75\textwidth,height=0.675\textwidth,xmin=-0.02,xmax=1.0,
                       ymin=0.0,ymax=1.02,legend pos=south east,grid=both,
                       xlabel={False Positive Rate},ylabel={True Positive Rate}] 
        \addplot[color=red,ultra thick] coordinates {
(1.000000,1.000000)
(0.980000,1.000000)
(0.960000,1.000000)
(0.940000,1.000000)
(0.920000,1.000000)
(0.900000,1.000000)
(0.880000,1.000000)
(0.860000,1.000000)
(0.840000,1.000000)
(0.820000,1.000000)
(0.800000,1.000000)
(0.780000,1.000000)
(0.760000,1.000000)
(0.740000,1.000000)
(0.720000,1.000000)
(0.700000,1.000000)
(0.680000,1.000000)
(0.680000,0.980000)
(0.680000,0.960000)
(0.680000,0.940000)
(0.680000,0.920000)
(0.680000,0.900000)
(0.660000,0.900000)
(0.660000,0.880000)
(0.640000,0.880000)
(0.620000,0.880000)
(0.600000,0.880000)
(0.580000,0.880000)
(0.560000,0.880000)
(0.560000,0.860000)
(0.560000,0.840000)
(0.540000,0.840000)
(0.540000,0.820000)
(0.540000,0.800000)
(0.540000,0.780000)
(0.540000,0.760000)
(0.520000,0.760000)
(0.500000,0.760000)
(0.480000,0.760000)
(0.460000,0.760000)
(0.440000,0.760000)
(0.440000,0.740000)
(0.420000,0.740000)
(0.420000,0.720000)
(0.420000,0.700000)
(0.420000,0.680000)
(0.400000,0.680000)
(0.400000,0.660000)
(0.380000,0.660000)
(0.360000,0.660000)
(0.340000,0.660000)
(0.320000,0.660000)
(0.300000,0.660000)
(0.280000,0.660000)
(0.260000,0.660000)
(0.260000,0.640000)
(0.260000,0.620000)
(0.260000,0.600000)
(0.260000,0.580000)
(0.260000,0.560000)
(0.260000,0.540000)
(0.260000,0.520000)
(0.260000,0.520000)
(0.260000,0.480000)
(0.260000,0.480000)
(0.260000,0.440000)
(0.260000,0.420000)
(0.260000,0.400000)
(0.260000,0.380000)
(0.260000,0.360000)
(0.260000,0.340000)
(0.260000,0.320000)
(0.260000,0.300000)
(0.260000,0.280000)
(0.260000,0.260000)
(0.240000,0.260000)
(0.240000,0.240000)
(0.240000,0.220000)
(0.240000,0.200000)
(0.220000,0.200000)
(0.200000,0.200000)
(0.180000,0.200000)
(0.180000,0.180000)
(0.180000,0.160000)
(0.160000,0.160000)
(0.160000,0.140000)
(0.160000,0.120000)
(0.160000,0.100000)
(0.140000,0.100000)
(0.120000,0.100000)
(0.100000,0.100000)
(0.080000,0.100000)
(0.060000,0.100000)
(0.040000,0.100000)
(0.020000,0.100000)
(0.000000,0.100000)
(0.000000,0.080000)
(0.000000,0.060000)
(0.000000,0.040000)
(0.000000,0.020000)
(0.000000,0.000000)
};
    \end{axis}
    \end{tikzpicture}
&
    \begin{tikzpicture}[scale=0.55]
    \begin{axis}[width=0.75\textwidth,height=0.675\textwidth,xmin=-0.02,xmax=1.0,
                       ymin=0.0,ymax=1.02,legend pos=south east,grid=both,
                       xlabel={False Positive Rate},ylabel={True Positive Rate}] 
        \addplot[color=red,ultra thick] coordinates {
(1.000000,1.000000)
(0.980000,1.000000)
(0.960000,1.000000)
(0.940000,1.000000)
(0.920000,1.000000)
(0.900000,1.000000)
(0.880000,1.000000)
(0.860000,1.000000)
(0.840000,1.000000)
(0.820000,1.000000)
(0.800000,1.000000)
(0.780000,1.000000)
(0.760000,1.000000)
(0.740000,1.000000)
(0.720000,1.000000)
(0.700000,1.000000)
(0.680000,1.000000)
(0.660000,1.000000)
(0.640000,1.000000)
(0.620000,1.000000)
(0.600000,1.000000)
(0.580000,1.000000)
(0.560000,1.000000)
(0.540000,1.000000)
(0.520000,1.000000)
(0.500000,1.000000)
(0.480000,1.000000)
(0.460000,1.000000)
(0.440000,1.000000)
(0.420000,1.000000)
(0.400000,1.000000)
(0.380000,1.000000)
(0.360000,1.000000)
(0.340000,1.000000)
(0.320000,1.000000)
(0.300000,1.000000)
(0.280000,1.000000)
(0.260000,1.000000)
(0.240000,1.000000)
(0.220000,1.000000)
(0.200000,1.000000)
(0.180000,1.000000)
(0.160000,1.000000)
(0.140000,1.000000)
(0.120000,1.000000)
(0.100000,1.000000)
(0.080000,1.000000)
(0.060000,1.000000)
(0.040000,1.000000)
(0.020000,1.000000)
(0.000000,1.000000)
(0.000000,0.980000)
(0.000000,0.960000)
(0.000000,0.940000)
(0.000000,0.920000)
(0.000000,0.900000)
(0.000000,0.880000)
(0.000000,0.860000)
(0.000000,0.840000)
(0.000000,0.820000)
(0.000000,0.800000)
(0.000000,0.780000)
(0.000000,0.760000)
(0.000000,0.740000)
(0.000000,0.720000)
(0.000000,0.700000)
(0.000000,0.680000)
(0.000000,0.660000)
(0.000000,0.640000)
(0.000000,0.620000)
(0.000000,0.600000)
(0.000000,0.580000)
(0.000000,0.560000)
(0.000000,0.540000)
(0.000000,0.520000)
(0.000000,0.500000)
(0.000000,0.480000)
(0.000000,0.460000)
(0.000000,0.440000)
(0.000000,0.420000)
(0.000000,0.400000)
(0.000000,0.380000)
(0.000000,0.360000)
(0.000000,0.340000)
(0.000000,0.320000)
(0.000000,0.300000)
(0.000000,0.280000)
(0.000000,0.260000)
(0.000000,0.240000)
(0.000000,0.220000)
(0.000000,0.200000)
(0.000000,0.180000)
(0.000000,0.160000)
(0.000000,0.140000)
(0.000000,0.120000)
(0.000000,0.100000)
(0.000000,0.080000)
(0.000000,0.060000)
(0.000000,0.040000)
(0.000000,0.020000)
(0.000000,0.000000)
};
    \end{axis}
    \end{tikzpicture}
\\
(c)  ROC Curve for Static Birthmarks & (d) ROC Curve for Dynamic Birthmarks
\\[2ex]
\end{tabular}
\end{center}
\caption{Security Shield HMM Results\label{fig:security_shield}}
\end{figure*}

\subsection{Results}

After performing each experiment, we plot the 
Receiver Operating Characteristic (ROC) curve and compute
area under the ROC curve (AUC). 
The ROC curve is obtained from a scatterplot by graphing the
true positive rate (TPR) versus the false positive rate (FPR)
as the threshold varies through the range of possible values.
The AUC can be interpreted as the 
probability that a randomly-selected positive instance scores
higher than a randomly selected negative instance~\cite{Bradley}.
Consequently, an AUC of~1.0 represents the ideal case
where a threshold exists that results in no classification errors.
On the other hand, an AUC of~0.5 implies that the binary
classifier is not better than flipping a coin.

We first perform our analysis on seven malware families using 
both static and dynamic analysis
techniques based on HMMs.
We then plot the ROC for each case and calculate the AUC.
To perform a comparison on both the techniques,
we have chosen the same data set in both cases.
We now consider a single family and examine the results.

For the Security Shield family,
Figure~\ref{fig:security_shield}~(a) shows the scatterplot
for the static HMM based on extracted opcode sequences.
The corresponding ROC curve appears in Figure~\ref{fig:security_shield}~(c).
The ROC curve in Figure~\ref{fig:security_shield}~(c)
yields an AUC of~0.676.

Again for the Security Shield family, the scatterplot for
the dynamic HMM based on extracted API call sequences
is given in Figure~\ref{fig:security_shield}~(b).
The corresponding ROC curve appears in Figure~\ref{fig:security_shield}~(d),
which has an AUC of~1.0.

HMMs based on static opcode sequences
have performed well when tested on various malware 
families~\cite{annachhatre,mwor,wongStamp}.
The AUC of~0.676 for the static HMM indicates that the Security Shield
represents a challenging detection problem
This makes the AUC of~1.0 for the dynamic
HMM all the more impressive.

We have performed experiments on each of the seven malware families
discussed in Section~\ref{sec:dataset}. For each family, we have tested
an HMM based on static opcodes and both an HMM and PHMM 
based on dynamic API calls. The results for the static and
dynamic HMMs are summarized in Table~\ref{tab:19}.
The ROC curves for the six families other than Security Shield
are given in the Appendix.

\begin{table}[htbp]
{\normalsize
\caption{AUC Curve Results for Static and Dynamic HMMs\label{tab:19}}
\begin{center}
\begin{tabular}{c|cc}\hline\hline
 & \multicolumn{2}{c}{AUC} \\
Malware family & \multicolumn{1}{c}{Dynamic} & \multicolumn{1}{c}{Static} \\ \hline
Cridex & 0.964	& 0.596  \\
Harebot & 0.974	& 0.622  \\
Security Shield & 1.000	& 0.676 \\
SmartHdd & 0.980	& 0.996  \\ 
Winwebsec & 0.985	& 0.835  \\
Zbot & 0.990 & 0.847  \\
Zeroaccess & 0.979	& 0.923  \\
\hline\hline
\end{tabular}
\end{center}
}
\end{table}

From the results in Table~\ref{tab:19} we see that the dynamic HMM
outperforms the static HMM in every case, except for the Smart HDD family,
in which case both the static and dynamic HMMs perform very well.
The average AUC for the dynamic HMM cases is~0.976,
whereas the average AUC for the static HMM cases is~0.785.
This clearly shows the advantage of dynamic birthmarks.

Next, we consider PHMMs trained on the same dynamic birthmarks
used for the dynamic HMMs. For a PHMM, the number of sequences 
used to train is a critical parameter.
Table~\ref{tab:phmm_auc2} contains results for our PHMM
experiments, where ``group~$n$'' means that we trained the corresponding
PHMM using~$n$ sequences. In each case, we were able to achieve
an AUC of~1.0, and once we have attained such a result, there is
no need for further experimentation with the parameter~$n$.

\begin{table*}[tp]
{\normalsize
\caption{PHMM Results\label{tab:phmm_auc2}}
\begin{center}
\begin{tabular}{c|ccccc}\hline\hline
\multirow{2}{*}{Malware Family}            & \multicolumn{5}{c}{AUC Results} \\
           & group 5 & group 10 & group 15 & group 20 & group 30 \\
  \hline
  \hline
Cridex & 0.958 & 1.000 & --- & --- & --- \\
Harebot & 0.875 & 0.952 & 1.000 & --- & --- \\
Security Shield & 0.988 & 0.964 & 1.000 & --- & --- \\
SmartHdd & 0.812 & 0.905 & 0.963 & 1.000 & --- \\
Winwebsec  & 0.997 & 0.995 & 1.000 & --- & --- \\
Zbot & 0.915 & 0.970 & 1.000 & --- & --- \\
Zeroaccess & 0.905 & 0.988 & 0.968 & 0.975 & 1.000 \\
\hline\hline
\end{tabular}
\end{center}
}
\end{table*}

The PHMM results in Table~\ref{tab:phmm_auc2} may be somewhat surprising
when compared to PHMM results obtained in previous research. For example,
in the paper~\cite{bib28}, a PHMM is trained on static opcode sequences
from metamorphic malware families. The PHMM results in~\cite{bib28}
are significantly worse than those obtained from HMMs trained on the same
static opcode data. Apparently, this is due to obfuscations that cause
opcode sequences to be shifted to different locations within the binaries,
causing gaps to proliferate when constructing the PHMMs, and thereby weakening the models. 
In contrast, for the dynamic API calls considered in this research 
it is far more difficult to obfuscate the sequential information. That is, the positional information
in API call sequences is highly informative, whereas the positional information
in opcode sequences is much less so.

Finally, in Figure~\label{fig:hmm_phmm_compare} we compare the results for our
static HMM, dynamic HMM, and dynamic PHMM, in the form of a bar graph.
While the dynamic HMM results are indeed very strong, the PHMM result
cannot be improved. 

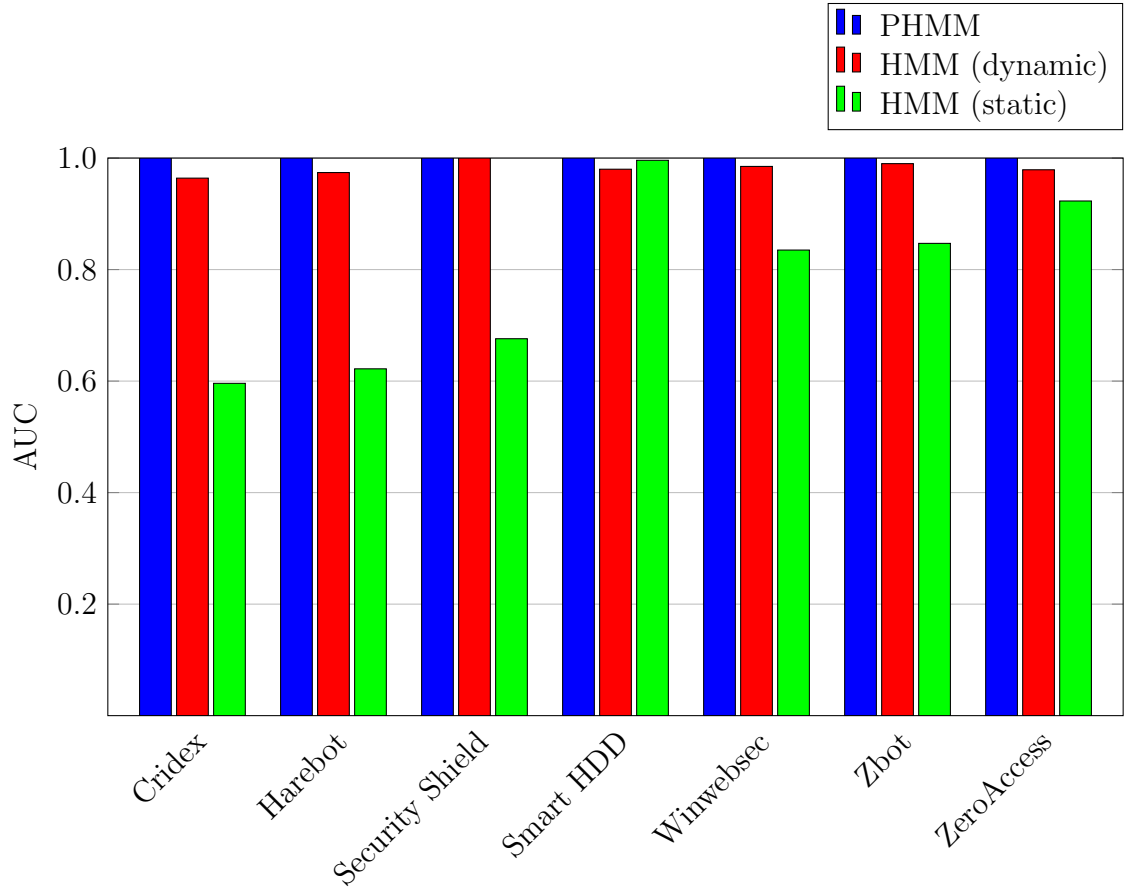
\begin{figure*}[ht]
\begin{center}
\begin{tikzpicture}
    \begin{axis}[
        width  = 0.9*\textwidth,
        height = 9cm,
        major x tick style = transparent,
        ybar=5*\pgflinewidth,
        bar width=12pt,
        ymajorgrids = true,
        ylabel = {AUC},
        symbolic x coords={Cridex,Harebot,Security Shield,Smart HDD,Winwebsec,Zbot,ZeroAccess},
	y tick label style={
    	/pgf/number format/.cd,
   	fixed,
   	fixed zerofill,
    	precision=1},
	ymin=0.0, ymax=1.0,
	ytick={0.2,0.4,0.6,0.8,1.0},
        xtick = data,
        x tick label style={rotate=45,anchor=north east, inner sep=0mm},
        scaled y ticks = false,
        enlarge x limits=0.1,
        ymin=0,
        legend cell align=left,
        legend style={
                at={(1,1.05)},
                anchor=south east,
                column sep=1ex
        }
    ]
        \addplot[fill=blue]
            coordinates {
(Cridex, 1.0)
(Harebot, 1.0)
(Security Shield, 1.0)
(Smart HDD, 1.0)
(Winwebsec, 1.0)
(Zbot, 1.0)
(ZeroAccess, 1.0)
};
        \addplot[fill=red]
            coordinates {
(Cridex, 0.964)
(Harebot, 0.974)
(Security Shield, 1.000)
(Smart HDD, 0.980)
(Winwebsec, 0.985)
(Zbot, 0.990)
(ZeroAccess, 0.979)
};
        \addplot[fill=green]
            coordinates {
(Cridex, 0.596)
(Harebot, 0.622)
(Security Shield, 0.676)
(Smart HDD, 0.996)
(Winwebsec, 0.835)
(Zbot, 0.847)
(ZeroAccess, 0.923)
};
        \legend{PHMM, HMM (dynamic), HMM (static)}
    \end{axis}
\end{tikzpicture}
\end{center}
\vglue-0.3in
\caption{PHMM versus HMMs\label{fig:hmm_phmm_compare}}
\end{figure*}


\section{Conclusion and Future work\label{chap:Conclusion and Future work}}

As malware has become increasingly difficult to detect with traditional techniques,
new approaches are needed,
and machine learning based tools seem a promising avenue.
Hidden Markov models are one such technique that has been
used effectively for malware detection in previous work~\cite{annachhatre,duelingHmm,kalbhor,kazi,rana,mwor,wongStamp}.
In contrast, Profile Hidden Markov Models seem to have rarely been studied in
this context, with previous results being mixed, at best~\cite{bib28}.

In this paper, we have shown the efficacy of detection tools
using HMMs built from dynamic birthmarks; specifically API calls.
In fact, this dynamic approach far outperforms previous static analysis tools
that relied on HMMs based on opcodes as static birthmarks.
The average area under the ROC curve (AUC) that
we obtained for all malware families in our dataset using our dynamic HMMs 
was~0.976, while the average AUC using static HMMs was~0.785.
The static HMMs have shown strong results in several previous studies, which indicates
that our dataset presents a relatively challenging case. This makes the dynamic
HMM results all the more impressive.

We also experimented with Profile Hidden Markov Models. The PHMM can be viewed 
as a generalization of an HMM that take positional information into account.
Our PHMM experiments, based again on dynamic API call sequences,
outperform the HMMs in every case. In fact, 
the PHMMs achieved ideal separation in every case, with an AUC of~1.0. 
These results show that PHMMs can be a very powerful tool for certain
types of software analysis problems. 


For future work,
we intend to further explore different morphing and obfuscation techniques
and test how effective our techniques are against different strategies.
In particular, it would be interesting to try to defeat 
HMM---and especially PHMM---techniques that are
based on dynamic API calls. By doing so, we can
hope to strengthen these detection techniques.
Also, while the dynamic birthmarks considered in this paper are based on API calls,
other dynamic birthmarks might provide beneficial information.
We also note that a potential strategy could be to combine the API calls
with other types of dynamic birthmarks using, say, Support Vector Machines
in order to develop an even robust malware detection model.
Finally, it would be interesting to test these dynamic tools
in the situation where we want to monitor untrusted code at runtime.

%
%

%
%


\section*{Appendix}
Here, we provide additional results from our experiments.
Figure~\ref{fig:s_roc}~(a) through~(f)
give ROC curves of HMMs trained on static opcode sequences, 
for Cridex, Harebot, Smart HDD, 
Winwebsec, Zbot, and Zeroaccess, respectively.
The ROC curves of HMMs trained on dynamic API call sequences
are given in Figure~\ref{fig:d_roc}~(a) through~(f).
The corresponding results for Security Shield are given in 
Figure~\ref{fig:security_shield}, parts~(c) and~(d).

\clearpage

%
%

\begin{figure*}[tp]
\begin{center}
\begin{tabular}{cc}
\begin{tikzpicture}[scale=0.50]
\begin{axis}[width=0.75\textwidth,height=0.675\textwidth,xmin=-0.02,xmax=1.0,
                   ymin=0.0,ymax=1.02,legend pos=south east,grid=both,
                   xlabel={False Positive Rate},ylabel={True Positive Rate}] 
\addplot[color=red,ultra thick] coordinates {
(1.000000,1.000000)
(0.980000,1.000000)
(0.960000,1.000000)
(0.940000,1.000000)
(0.920000,1.000000)
(0.900000,1.000000)
(0.900000,0.980000)
(0.880000,0.980000)
(0.860000,0.980000)
(0.840000,0.980000)
(0.840000,0.960000)
(0.820000,0.960000)
(0.800000,0.960000)
(0.780000,0.960000)
(0.760000,0.960000)
(0.740000,0.960000)
(0.720000,0.960000)
(0.700000,0.960000)
(0.680000,0.960000)
(0.660000,0.960000)
(0.640000,0.960000)
(0.620000,0.960000)
(0.600000,0.960000)
(0.600000,0.940000)
(0.600000,0.920000)
(0.600000,0.900000)
(0.600000,0.880000)
(0.600000,0.860000)
(0.600000,0.840000)
(0.600000,0.820000)
(0.600000,0.800000)
(0.600000,0.780000)
(0.580000,0.780000)
(0.560000,0.780000)
(0.560000,0.760000)
(0.540000,0.760000)
(0.520000,0.760000)
(0.500000,0.760000)
(0.480000,0.760000)
(0.480000,0.740000)
(0.480000,0.720000)
(0.480000,0.700000)
(0.480000,0.680000)
(0.480000,0.680000)
(0.480000,0.640000)
(0.480000,0.620000)
(0.480000,0.600000)
(0.480000,0.580000)
(0.480000,0.560000)
(0.480000,0.540000)
(0.480000,0.520000)
(0.480000,0.500000)
(0.480000,0.480000)
(0.480000,0.460000)
(0.480000,0.440000)
(0.480000,0.420000)
(0.480000,0.400000)
(0.480000,0.380000)
(0.460000,0.380000)
(0.460000,0.360000)
(0.440000,0.360000)
(0.440000,0.340000)
(0.440000,0.320000)
(0.420000,0.320000)
(0.400000,0.320000)
(0.380000,0.320000)
(0.380000,0.300000)
(0.360000,0.300000)
(0.360000,0.280000)
(0.340000,0.280000)
(0.320000,0.280000)
(0.300000,0.280000)
(0.280000,0.280000)
(0.260000,0.280000)
(0.240000,0.280000)
(0.240000,0.260000)
(0.240000,0.240000)
(0.220000,0.240000)
(0.200000,0.240000)
(0.200000,0.220000)
(0.180000,0.220000)
(0.180000,0.200000)
(0.180000,0.180000)
(0.160000,0.180000)
(0.160000,0.160000)
(0.140000,0.160000)
(0.120000,0.160000)
(0.100000,0.160000)
(0.080000,0.160000)
(0.060000,0.160000)
(0.040000,0.160000)
(0.040000,0.140000)
(0.020000,0.140000)
(0.020000,0.120000)
(0.020000,0.100000)
(0.000000,0.100000)
(0.000000,0.080000)
(0.000000,0.060000)
(0.000000,0.040000)
(0.000000,0.020000)
(0.000000,0.000000)
};
\end{axis}
\end{tikzpicture}
&
\begin{tikzpicture}[scale=0.50]
\begin{axis}[width=0.75\textwidth,height=0.675\textwidth,xmin=-0.02,xmax=1.0,
                   ymin=0.0,ymax=1.02,legend pos=south east,grid=both,
                   xlabel={False Positive Rate},ylabel={True Positive Rate}] 
\addplot[color=red,ultra thick] coordinates {
(1.000000,1.000000)
(0.980000,1.000000)
(0.960000,1.000000)
(0.940000,1.000000)
(0.920000,1.000000)
(0.900000,1.000000)
(0.880000,1.000000)
(0.860000,1.000000)
(0.840000,1.000000)
(0.820000,1.000000)
(0.800000,1.000000)
(0.780000,1.000000)
(0.760000,1.000000)
(0.740000,1.000000)
(0.720000,1.000000)
(0.700000,1.000000)
(0.680000,1.000000)
(0.660000,1.000000)
(0.640000,1.000000)
(0.620000,1.000000)
(0.600000,1.000000)
(0.600000,1.000000)
(0.600000,0.960000)
(0.600000,0.940000)
(0.600000,0.920000)
(0.600000,0.900000)
(0.600000,0.880000)
(0.600000,0.860000)
(0.600000,0.840000)
(0.600000,0.820000)
(0.600000,0.800000)
(0.600000,0.780000)
(0.600000,0.760000)
(0.600000,0.740000)
(0.600000,0.720000)
(0.600000,0.700000)
(0.600000,0.680000)
(0.600000,0.660000)
(0.580000,0.660000)
(0.560000,0.660000)
(0.560000,0.640000)
(0.540000,0.640000)
(0.520000,0.640000)
(0.500000,0.640000)
(0.500000,0.620000)
(0.500000,0.600000)
(0.480000,0.600000)
(0.460000,0.600000)
(0.440000,0.600000)
(0.420000,0.600000)
(0.400000,0.600000)
(0.400000,0.580000)
(0.400000,0.560000)
(0.400000,0.540000)
(0.380000,0.540000)
(0.380000,0.520000)
(0.380000,0.500000)
(0.360000,0.500000)
(0.360000,0.480000)
(0.340000,0.480000)
(0.320000,0.480000)
(0.320000,0.460000)
(0.320000,0.440000)
(0.320000,0.420000)
(0.300000,0.420000)
(0.300000,0.400000)
(0.280000,0.400000)
(0.280000,0.380000)
(0.280000,0.360000)
(0.280000,0.340000)
(0.280000,0.320000)
(0.280000,0.300000)
(0.280000,0.280000)
(0.260000,0.280000)
(0.240000,0.280000)
(0.240000,0.260000)
(0.240000,0.240000)
(0.240000,0.220000)
(0.240000,0.200000)
(0.240000,0.180000)
(0.220000,0.180000)
(0.200000,0.180000)
(0.200000,0.160000)
(0.180000,0.160000)
(0.160000,0.160000)
(0.140000,0.160000)
(0.120000,0.160000)
(0.120000,0.140000)
(0.120000,0.120000)
(0.100000,0.120000)
(0.080000,0.120000)
(0.060000,0.120000)
(0.040000,0.120000)
(0.040000,0.100000)
(0.040000,0.080000)
(0.040000,0.060000)
(0.020000,0.060000)
(0.000000,0.060000)
(0.000000,0.040000)
(0.000000,0.020000)
(0.000000,0.000000)
};
\end{axis}
\end{tikzpicture}
\\
(a) Cridex & (b) Harebot
\\[2ex]
\begin{tikzpicture}[scale=0.50]
\begin{axis}[width=0.75\textwidth,height=0.675\textwidth,xmin=-0.02,xmax=1.0,
                   ymin=0.0,ymax=1.02,legend pos=south east,grid=both,
                   xlabel={False Positive Rate},ylabel={True Positive Rate}] 
\addplot[color=red,ultra thick] coordinates {
(1.000000,1.000000)
(0.980000,1.000000)
(0.960000,1.000000)
(0.940000,1.000000)
(0.920000,1.000000)
(0.900000,1.000000)
(0.880000,1.000000)
(0.860000,1.000000)
(0.840000,1.000000)
(0.820000,1.000000)
(0.800000,1.000000)
(0.780000,1.000000)
(0.760000,1.000000)
(0.740000,1.000000)
(0.720000,1.000000)
(0.700000,1.000000)
(0.680000,1.000000)
(0.660000,1.000000)
(0.640000,1.000000)
(0.620000,1.000000)
(0.600000,1.000000)
(0.580000,1.000000)
(0.560000,1.000000)
(0.540000,1.000000)
(0.520000,1.000000)
(0.500000,1.000000)
(0.480000,1.000000)
(0.460000,1.000000)
(0.440000,1.000000)
(0.420000,1.000000)
(0.400000,1.000000)
(0.380000,1.000000)
(0.360000,1.000000)
(0.340000,1.000000)
(0.320000,1.000000)
(0.300000,1.000000)
(0.280000,1.000000)
(0.260000,1.000000)
(0.240000,1.000000)
(0.220000,1.000000)
(0.200000,1.000000)
(0.200000,0.980000)
(0.180000,0.980000)
(0.160000,0.980000)
(0.140000,0.980000)
(0.120000,0.980000)
(0.100000,0.980000)
(0.080000,0.980000)
(0.060000,0.980000)
(0.040000,0.980000)
(0.020000,0.980000)
(0.000000,0.980000)
(0.000000,0.960000)
(0.000000,0.940000)
(0.000000,0.920000)
(0.000000,0.900000)
(0.000000,0.880000)
(0.000000,0.860000)
(0.000000,0.840000)
(0.000000,0.820000)
(0.000000,0.800000)
(0.000000,0.780000)
(0.000000,0.760000)
(0.000000,0.740000)
(0.000000,0.720000)
(0.000000,0.700000)
(0.000000,0.680000)
(0.000000,0.660000)
(0.000000,0.640000)
(0.000000,0.620000)
(0.000000,0.600000)
(0.000000,0.580000)
(0.000000,0.560000)
(0.000000,0.540000)
(0.000000,0.520000)
(0.000000,0.500000)
(0.000000,0.480000)
(0.000000,0.460000)
(0.000000,0.440000)
(0.000000,0.420000)
(0.000000,0.400000)
(0.000000,0.380000)
(0.000000,0.360000)
(0.000000,0.340000)
(0.000000,0.320000)
(0.000000,0.300000)
(0.000000,0.280000)
(0.000000,0.260000)
(0.000000,0.240000)
(0.000000,0.220000)
(0.000000,0.200000)
(0.000000,0.180000)
(0.000000,0.160000)
(0.000000,0.140000)
(0.000000,0.120000)
(0.000000,0.100000)
(0.000000,0.080000)
(0.000000,0.060000)
(0.000000,0.040000)
(0.000000,0.020000)
(0.000000,0.000000)
};
\end{axis}
\end{tikzpicture}
&
\begin{tikzpicture}[scale=0.50]
\begin{axis}[width=0.75\textwidth,height=0.675\textwidth,xmin=-0.02,xmax=1.0,
                   ymin=0.0,ymax=1.02,legend pos=south east,grid=both,
                   xlabel={False Positive Rate},ylabel={True Positive Rate}] 
\addplot[color=red,ultra thick] coordinates {
(1.000000,1.000000)
(0.990000,1.000000)
(0.980000,1.000000)
(0.970000,1.000000)
(0.960000,1.000000)
(0.950000,1.000000)
(0.940000,1.000000)
(0.930000,1.000000)
(0.920000,1.000000)
(0.910000,1.000000)
(0.910000,1.000000)
(0.910000,0.980000)
(0.900000,0.980000)
(0.890000,0.980000)
(0.880000,0.980000)
(0.870000,0.980000)
(0.860000,0.980000)
(0.860000,0.970000)
(0.850000,0.970000)
(0.850000,0.960000)
(0.840000,0.960000)
(0.840000,0.950000)
(0.830000,0.950000)
(0.820000,0.950000)
(0.810000,0.950000)
(0.800000,0.950000)
(0.790000,0.950000)
(0.780000,0.950000)
(0.770000,0.950000)
(0.760000,0.950000)
(0.750000,0.950000)
(0.740000,0.950000)
(0.730000,0.950000)
(0.720000,0.950000)
(0.710000,0.950000)
(0.700000,0.950000)
(0.690000,0.950000)
(0.680000,0.950000)
(0.670000,0.950000)
(0.660000,0.950000)
(0.650000,0.950000)
(0.640000,0.950000)
(0.630000,0.950000)
(0.620000,0.950000)
(0.610000,0.950000)
(0.600000,0.950000)
(0.590000,0.950000)
(0.580000,0.950000)
(0.570000,0.950000)
(0.560000,0.950000)
(0.550000,0.950000)
(0.540000,0.950000)
(0.530000,0.950000)
(0.520000,0.950000)
(0.510000,0.950000)
(0.500000,0.950000)
(0.490000,0.950000)
(0.480000,0.950000)
(0.470000,0.950000)
(0.460000,0.950000)
(0.460000,0.940000)
(0.450000,0.940000)
(0.440000,0.940000)
(0.430000,0.940000)
(0.420000,0.940000)
(0.410000,0.940000)
(0.400000,0.940000)
(0.390000,0.940000)
(0.380000,0.940000)
(0.370000,0.940000)
(0.360000,0.940000)
(0.350000,0.940000)
(0.350000,0.940000)
(0.350000,0.920000)
(0.340000,0.920000)
(0.330000,0.920000)
(0.320000,0.920000)
(0.320000,0.920000)
(0.300000,0.920000)
(0.300000,0.920000)
(0.280000,0.920000)
(0.280000,0.910000)
(0.280000,0.900000)
(0.280000,0.890000)
(0.280000,0.890000)
(0.260000,0.890000)
(0.260000,0.890000)
(0.240000,0.890000)
(0.230000,0.890000)
(0.230000,0.890000)
(0.210000,0.890000)
(0.200000,0.890000)
(0.200000,0.880000)
(0.200000,0.870000)
(0.200000,0.860000)
(0.200000,0.860000)
(0.200000,0.840000)
(0.200000,0.830000)
(0.200000,0.820000)
(0.200000,0.810000)
(0.200000,0.800000)
(0.200000,0.790000)
(0.200000,0.780000)
(0.200000,0.770000)
(0.200000,0.760000)
(0.200000,0.750000)
(0.200000,0.740000)
(0.200000,0.730000)
(0.200000,0.720000)
(0.190000,0.720000)
(0.190000,0.710000)
(0.190000,0.700000)
(0.190000,0.690000)
(0.190000,0.680000)
(0.190000,0.670000)
(0.190000,0.670000)
(0.190000,0.650000)
(0.190000,0.640000)
(0.190000,0.630000)
(0.190000,0.620000)
(0.190000,0.610000)
(0.190000,0.600000)
(0.190000,0.590000)
(0.190000,0.580000)
(0.190000,0.570000)
(0.190000,0.560000)
(0.190000,0.560000)
(0.190000,0.540000)
(0.190000,0.530000)
(0.180000,0.530000)
(0.180000,0.520000)
(0.180000,0.510000)
(0.180000,0.500000)
(0.170000,0.500000)
(0.160000,0.500000)
(0.150000,0.500000)
(0.140000,0.500000)
(0.130000,0.500000)
(0.120000,0.500000)
(0.120000,0.490000)
(0.110000,0.490000)
(0.100000,0.490000)
(0.100000,0.480000)
(0.090000,0.480000)
(0.080000,0.480000)
(0.080000,0.470000)
(0.070000,0.470000)
(0.070000,0.460000)
(0.070000,0.450000)
(0.070000,0.440000)
(0.070000,0.430000)
(0.070000,0.420000)
(0.070000,0.410000)
(0.060000,0.410000)
(0.060000,0.400000)
(0.060000,0.390000)
(0.060000,0.380000)
(0.060000,0.370000)
(0.060000,0.360000)
(0.060000,0.350000)
(0.060000,0.340000)
(0.060000,0.330000)
(0.060000,0.320000)
(0.060000,0.310000)
(0.060000,0.300000)
(0.060000,0.290000)
(0.060000,0.280000)
(0.060000,0.270000)
(0.060000,0.260000)
(0.060000,0.250000)
(0.060000,0.240000)
(0.060000,0.230000)
(0.060000,0.220000)
(0.060000,0.210000)
(0.060000,0.200000)
(0.060000,0.190000)
(0.060000,0.180000)
(0.060000,0.170000)
(0.060000,0.160000)
(0.060000,0.150000)
(0.060000,0.140000)
(0.060000,0.130000)
(0.060000,0.120000)
(0.050000,0.120000)
(0.040000,0.120000)
(0.030000,0.120000)
(0.020000,0.120000)
(0.020000,0.110000)
(0.010000,0.110000)
(0.010000,0.110000)
(0.010000,0.090000)
(0.000000,0.090000)
(0.000000,0.080000)
(0.000000,0.070000)
(0.000000,0.060000)
(0.000000,0.050000)
(0.000000,0.040000)
(0.000000,0.030000)
(0.000000,0.020000)
(0.000000,0.010000)
(0.000000,0.000000)
};
\end{axis}
\end{tikzpicture}
\\
(c) Smart HDD & (d) Winwebsec
\\[2ex]
\begin{tikzpicture}[scale=0.50]
\begin{axis}[width=0.75\textwidth,height=0.675\textwidth,xmin=-0.02,xmax=1.0,
                   ymin=0.0,ymax=1.02,legend pos=south east,grid=both,
                   xlabel={False Positive Rate},ylabel={True Positive Rate}] 
\addplot[color=red,ultra thick] coordinates {
(1.000000,1.000000)
(0.990000,1.000000)
(0.980000,1.000000)
(0.970000,1.000000)
(0.960000,1.000000)
(0.950000,1.000000)
(0.940000,1.000000)
(0.930000,1.000000)
(0.920000,1.000000)
(0.910000,1.000000)
(0.900000,1.000000)
(0.890000,1.000000)
(0.880000,1.000000)
(0.870000,1.000000)
(0.860000,1.000000)
(0.850000,1.000000)
(0.840000,1.000000)
(0.830000,1.000000)
(0.820000,1.000000)
(0.810000,1.000000)
(0.800000,1.000000)
(0.790000,1.000000)
(0.780000,1.000000)
(0.770000,1.000000)
(0.760000,1.000000)
(0.750000,1.000000)
(0.740000,1.000000)
(0.730000,1.000000)
(0.720000,1.000000)
(0.710000,1.000000)
(0.700000,1.000000)
(0.690000,1.000000)
(0.680000,1.000000)
(0.670000,1.000000)
(0.660000,1.000000)
(0.650000,1.000000)
(0.640000,1.000000)
(0.630000,1.000000)
(0.620000,1.000000)
(0.610000,1.000000)
(0.600000,1.000000)
(0.600000,0.990000)
(0.590000,0.990000)
(0.580000,0.990000)
(0.570000,0.990000)
(0.560000,0.990000)
(0.550000,0.990000)
(0.540000,0.990000)
(0.530000,0.990000)
(0.530000,0.980000)
(0.520000,0.980000)
(0.520000,0.970000)
(0.520000,0.960000)
(0.510000,0.960000)
(0.510000,0.950000)
(0.500000,0.950000)
(0.500000,0.940000)
(0.500000,0.930000)
(0.500000,0.920000)
(0.500000,0.910000)
(0.490000,0.910000)
(0.480000,0.910000)
(0.470000,0.910000)
(0.460000,0.910000)
(0.450000,0.910000)
(0.440000,0.910000)
(0.430000,0.910000)
(0.420000,0.910000)
(0.410000,0.910000)
(0.410000,0.900000)
(0.400000,0.900000)
(0.390000,0.900000)
(0.380000,0.900000)
(0.370000,0.900000)
(0.370000,0.890000)
(0.360000,0.890000)
(0.350000,0.890000)
(0.340000,0.890000)
(0.330000,0.890000)
(0.330000,0.880000)
(0.330000,0.870000)
(0.320000,0.870000)
(0.310000,0.870000)
(0.300000,0.870000)
(0.300000,0.860000)
(0.290000,0.860000)
(0.280000,0.860000)
(0.270000,0.860000)
(0.270000,0.850000)
(0.260000,0.850000)
(0.260000,0.840000)
(0.250000,0.840000)
(0.250000,0.830000)
(0.250000,0.820000)
(0.250000,0.810000)
(0.250000,0.800000)
(0.250000,0.790000)
(0.250000,0.780000)
(0.240000,0.780000)
(0.230000,0.780000)
(0.230000,0.770000)
(0.230000,0.760000)
(0.230000,0.750000)
(0.230000,0.740000)
(0.230000,0.730000)
(0.230000,0.720000)
(0.220000,0.720000)
(0.220000,0.710000)
(0.210000,0.710000)
(0.210000,0.700000)
(0.210000,0.690000)
(0.210000,0.680000)
(0.210000,0.670000)
(0.210000,0.660000)
(0.210000,0.650000)
(0.210000,0.640000)
(0.210000,0.630000)
(0.210000,0.620000)
(0.210000,0.610000)
(0.210000,0.600000)
(0.210000,0.590000)
(0.210000,0.580000)
(0.200000,0.580000)
(0.200000,0.570000)
(0.200000,0.560000)
(0.200000,0.550000)
(0.200000,0.550000)
(0.200000,0.530000)
(0.200000,0.520000)
(0.190000,0.520000)
(0.190000,0.510000)
(0.180000,0.510000)
(0.180000,0.500000)
(0.180000,0.490000)
(0.180000,0.480000)
(0.170000,0.480000)
(0.160000,0.480000)
(0.160000,0.470000)
(0.160000,0.460000)
(0.150000,0.460000)
(0.140000,0.460000)
(0.130000,0.460000)
(0.130000,0.460000)
(0.110000,0.460000)
(0.100000,0.460000)
(0.100000,0.460000)
(0.080000,0.460000)
(0.070000,0.460000)
(0.070000,0.450000)
(0.060000,0.450000)
(0.060000,0.450000)
(0.040000,0.450000)
(0.040000,0.450000)
(0.020000,0.450000)
(0.020000,0.440000)
(0.020000,0.430000)
(0.020000,0.420000)
(0.020000,0.410000)
(0.020000,0.410000)
(0.020000,0.390000)
(0.020000,0.380000)
(0.020000,0.370000)
(0.020000,0.360000)
(0.020000,0.350000)
(0.020000,0.350000)
(0.000000,0.350000)
(0.000000,0.340000)
(0.000000,0.340000)
(0.000000,0.320000)
(0.000000,0.310000)
(0.000000,0.300000)
(0.000000,0.290000)
(0.000000,0.280000)
(0.000000,0.270000)
(0.000000,0.260000)
(0.000000,0.260000)
(0.000000,0.240000)
(0.000000,0.230000)
(0.000000,0.230000)
(0.000000,0.210000)
(0.000000,0.200000)
(0.000000,0.190000)
(0.000000,0.180000)
(0.000000,0.170000)
(0.000000,0.160000)
(0.000000,0.150000)
(0.000000,0.140000)
(0.000000,0.130000)
(0.000000,0.120000)
(0.000000,0.110000)
(0.000000,0.100000)
(0.000000,0.090000)
(0.000000,0.080000)
(0.000000,0.070000)
(0.000000,0.060000)
(0.000000,0.050000)
(0.000000,0.040000)
(0.000000,0.030000)
(0.000000,0.020000)
(0.000000,0.020000)
(0.000000,0.000000)
};
\end{axis}
\end{tikzpicture}
&
\begin{tikzpicture}[scale=0.50]
\begin{axis}[width=0.75\textwidth,height=0.675\textwidth,xmin=-0.02,xmax=1.0,
                   ymin=0.0,ymax=1.02,legend pos=south east,grid=both,
                   xlabel={False Positive Rate},ylabel={True Positive Rate}] 
\addplot[color=red,ultra thick] coordinates {
(1.000000,1.000000)
(0.990000,1.000000)
(0.980000,1.000000)
(0.970000,1.000000)
(0.960000,1.000000)
(0.950000,1.000000)
(0.940000,1.000000)
(0.930000,1.000000)
(0.920000,1.000000)
(0.910000,1.000000)
(0.900000,1.000000)
(0.890000,1.000000)
(0.880000,1.000000)
(0.870000,1.000000)
(0.860000,1.000000)
(0.850000,1.000000)
(0.840000,1.000000)
(0.830000,1.000000)
(0.820000,1.000000)
(0.810000,1.000000)
(0.800000,1.000000)
(0.790000,1.000000)
(0.780000,1.000000)
(0.770000,1.000000)
(0.760000,1.000000)
(0.750000,1.000000)
(0.740000,1.000000)
(0.730000,1.000000)
(0.720000,1.000000)
(0.710000,1.000000)
(0.700000,1.000000)
(0.690000,1.000000)
(0.680000,1.000000)
(0.670000,1.000000)
(0.660000,1.000000)
(0.650000,1.000000)
(0.640000,1.000000)
(0.630000,1.000000)
(0.620000,1.000000)
(0.610000,1.000000)
(0.600000,1.000000)
(0.590000,1.000000)
(0.580000,1.000000)
(0.580000,0.990000)
(0.570000,0.990000)
(0.560000,0.990000)
(0.550000,0.990000)
(0.540000,0.990000)
(0.540000,0.980000)
(0.530000,0.980000)
(0.520000,0.980000)
(0.510000,0.980000)
(0.500000,0.980000)
(0.490000,0.980000)
(0.480000,0.980000)
(0.470000,0.980000)
(0.460000,0.980000)
(0.460000,0.970000)
(0.450000,0.970000)
(0.440000,0.970000)
(0.430000,0.970000)
(0.420000,0.970000)
(0.410000,0.970000)
(0.400000,0.970000)
(0.390000,0.970000)
(0.380000,0.970000)
(0.370000,0.970000)
(0.360000,0.970000)
(0.350000,0.970000)
(0.340000,0.970000)
(0.330000,0.970000)
(0.330000,0.960000)
(0.320000,0.960000)
(0.310000,0.960000)
(0.310000,0.950000)
(0.300000,0.950000)
(0.290000,0.950000)
(0.280000,0.950000)
(0.270000,0.950000)
(0.270000,0.940000)
(0.260000,0.940000)
(0.250000,0.940000)
(0.250000,0.930000)
(0.250000,0.930000)
(0.250000,0.910000)
(0.240000,0.910000)
(0.230000,0.910000)
(0.230000,0.900000)
(0.230000,0.890000)
(0.230000,0.880000)
(0.230000,0.870000)
(0.230000,0.860000)
(0.230000,0.850000)
(0.220000,0.850000)
(0.220000,0.840000)
(0.210000,0.840000)
(0.200000,0.840000)
(0.200000,0.830000)
(0.200000,0.820000)
(0.200000,0.810000)
(0.200000,0.800000)
(0.200000,0.790000)
(0.190000,0.790000)
(0.180000,0.790000)
(0.170000,0.790000)
(0.170000,0.780000)
(0.160000,0.780000)
(0.160000,0.780000)
(0.160000,0.760000)
(0.160000,0.750000)
(0.160000,0.740000)
(0.150000,0.740000)
(0.140000,0.740000)
(0.140000,0.740000)
(0.120000,0.740000)
(0.110000,0.740000)
(0.100000,0.740000)
(0.090000,0.740000)
(0.090000,0.740000)
(0.070000,0.740000)
(0.060000,0.740000)
(0.060000,0.740000)
(0.060000,0.720000)
(0.060000,0.710000)
(0.060000,0.700000)
(0.060000,0.690000)
(0.060000,0.680000)
(0.060000,0.670000)
(0.060000,0.660000)
(0.060000,0.660000)
(0.060000,0.640000)
(0.060000,0.630000)
(0.060000,0.620000)
(0.060000,0.610000)
(0.060000,0.600000)
(0.060000,0.590000)
(0.060000,0.580000)
(0.060000,0.570000)
(0.060000,0.570000)
(0.040000,0.570000)
(0.040000,0.570000)
(0.020000,0.570000)
(0.020000,0.570000)
(0.000000,0.570000)
(0.000000,0.560000)
(0.000000,0.550000)
(0.000000,0.540000)
(0.000000,0.530000)
(0.000000,0.520000)
(0.000000,0.510000)
(0.000000,0.500000)
(0.000000,0.490000)
(0.000000,0.480000)
(0.000000,0.470000)
(0.000000,0.460000)
(0.000000,0.450000)
(0.000000,0.440000)
(0.000000,0.430000)
(0.000000,0.420000)
(0.000000,0.410000)
(0.000000,0.400000)
(0.000000,0.390000)
(0.000000,0.380000)
(0.000000,0.370000)
(0.000000,0.360000)
(0.000000,0.350000)
(0.000000,0.340000)
(0.000000,0.330000)
(0.000000,0.320000)
(0.000000,0.310000)
(0.000000,0.300000)
(0.000000,0.290000)
(0.000000,0.280000)
(0.000000,0.270000)
(0.000000,0.260000)
(0.000000,0.250000)
(0.000000,0.250000)
(0.000000,0.250000)
(0.000000,0.220000)
(0.000000,0.220000)
(0.000000,0.200000)
(0.000000,0.190000)
(0.000000,0.180000)
(0.000000,0.170000)
(0.000000,0.160000)
(0.000000,0.150000)
(0.000000,0.140000)
(0.000000,0.130000)
(0.000000,0.120000)
(0.000000,0.110000)
(0.000000,0.100000)
(0.000000,0.090000)
(0.000000,0.080000)
(0.000000,0.070000)
(0.000000,0.060000)
(0.000000,0.050000)
(0.000000,0.040000)
(0.000000,0.030000)
(0.000000,0.020000)
(0.000000,0.010000)
(0.000000,0.000000)
};
\end{axis}
\end{tikzpicture}
\\
(e) Zbot & (f) Zeroaccess
\\[2ex]
\end{tabular}
\end{center}
\vglue -0.20in
\caption{ROC Curves for HMMs Based on Static Birthmarks\label{fig:s_roc}}
\end{figure*}

\clearpage

%
%

\begin{figure*}[tp]
\begin{center}
\begin{tabular}{cc}
\begin{tikzpicture}[scale=0.50]
\begin{axis}[width=0.75\textwidth,height=0.675\textwidth,xmin=-0.02,xmax=1.0,
                   ymin=0.0,ymax=1.02,legend pos=south east,grid=both,
                   xlabel={False Positive Rate},ylabel={True Positive Rate}] 
\addplot[color=red,ultra thick] coordinates {
(1.000000,1.000000)
(1.000000,0.980000)
(0.980000,0.980000)
(0.960000,0.980000)
(0.940000,0.980000)
(0.920000,0.980000)
(0.900000,0.980000)
(0.880000,0.980000)
(0.860000,0.980000)
(0.840000,0.980000)
(0.820000,0.980000)
(0.800000,0.980000)
(0.780000,0.980000)
(0.760000,0.980000)
(0.740000,0.980000)
(0.720000,0.980000)
(0.700000,0.980000)
(0.680000,0.980000)
(0.660000,0.980000)
(0.640000,0.980000)
(0.620000,0.980000)
(0.600000,0.980000)
(0.580000,0.980000)
(0.560000,0.980000)
(0.540000,0.980000)
(0.520000,0.980000)
(0.500000,0.980000)
(0.480000,0.980000)
(0.460000,0.980000)
(0.440000,0.980000)
(0.420000,0.980000)
(0.400000,0.980000)
(0.380000,0.980000)
(0.360000,0.980000)
(0.340000,0.980000)
(0.320000,0.980000)
(0.300000,0.980000)
(0.280000,0.980000)
(0.280000,0.960000)
(0.260000,0.960000)
(0.240000,0.960000)
(0.220000,0.960000)
(0.200000,0.960000)
(0.180000,0.960000)
(0.160000,0.960000)
(0.140000,0.960000)
(0.120000,0.960000)
(0.100000,0.960000)
(0.080000,0.960000)
(0.060000,0.960000)
(0.040000,0.960000)
(0.020000,0.960000)
(0.020000,0.940000)
(0.020000,0.920000)
(0.020000,0.900000)
(0.020000,0.880000)
(0.020000,0.860000)
(0.020000,0.840000)
(0.020000,0.820000)
(0.020000,0.800000)
(0.020000,0.780000)
(0.020000,0.760000)
(0.020000,0.740000)
(0.020000,0.720000)
(0.020000,0.700000)
(0.020000,0.680000)
(0.020000,0.660000)
(0.020000,0.640000)
(0.020000,0.620000)
(0.020000,0.600000)
(0.020000,0.580000)
(0.020000,0.560000)
(0.020000,0.540000)
(0.020000,0.520000)
(0.020000,0.500000)
(0.020000,0.480000)
(0.020000,0.460000)
(0.000000,0.460000)
(0.000000,0.440000)
(0.000000,0.420000)
(0.000000,0.400000)
(0.000000,0.380000)
(0.000000,0.360000)
(0.000000,0.340000)
(0.000000,0.320000)
(0.000000,0.300000)
(0.000000,0.280000)
(0.000000,0.260000)
(0.000000,0.240000)
(0.000000,0.220000)
(0.000000,0.200000)
(0.000000,0.180000)
(0.000000,0.160000)
(0.000000,0.140000)
(0.000000,0.120000)
(0.000000,0.100000)
(0.000000,0.080000)
(0.000000,0.060000)
(0.000000,0.040000)
(0.000000,0.020000)
(0.000000,0.000000)
};
\end{axis}
\end{tikzpicture}
&
\begin{tikzpicture}[scale=0.50]
\begin{axis}[width=0.75\textwidth,height=0.675\textwidth,xmin=-0.02,xmax=1.0,
                   ymin=0.0,ymax=1.02,legend pos=south east,grid=both,
                   xlabel={False Positive Rate},ylabel={True Positive Rate}] 
\addplot[color=red,ultra thick] coordinates {
(1.000000,1.000000)
(1.000000,0.980000)
(0.980000,0.980000)
(0.960000,0.980000)
(0.940000,0.980000)
(0.920000,0.980000)
(0.900000,0.980000)
(0.880000,0.980000)
(0.860000,0.980000)
(0.840000,0.980000)
(0.820000,0.980000)
(0.800000,0.980000)
(0.780000,0.980000)
(0.760000,0.980000)
(0.740000,0.980000)
(0.720000,0.980000)
(0.700000,0.980000)
(0.680000,0.980000)
(0.660000,0.980000)
(0.640000,0.980000)
(0.620000,0.980000)
(0.600000,0.980000)
(0.580000,0.980000)
(0.560000,0.980000)
(0.540000,0.980000)
(0.520000,0.980000)
(0.500000,0.980000)
(0.480000,0.980000)
(0.460000,0.980000)
(0.440000,0.980000)
(0.420000,0.980000)
(0.400000,0.980000)
(0.380000,0.980000)
(0.360000,0.980000)
(0.340000,0.980000)
(0.320000,0.980000)
(0.300000,0.980000)
(0.280000,0.980000)
(0.260000,0.980000)
(0.260000,0.960000)
(0.240000,0.960000)
(0.220000,0.960000)
(0.200000,0.960000)
(0.180000,0.960000)
(0.160000,0.960000)
(0.140000,0.960000)
(0.120000,0.960000)
(0.100000,0.960000)
(0.080000,0.960000)
(0.060000,0.960000)
(0.040000,0.960000)
(0.020000,0.960000)
(0.000000,0.960000)
(0.000000,0.940000)
(0.000000,0.920000)
(0.000000,0.900000)
(0.000000,0.880000)
(0.000000,0.860000)
(0.000000,0.840000)
(0.000000,0.820000)
(0.000000,0.800000)
(0.000000,0.780000)
(0.000000,0.760000)
(0.000000,0.740000)
(0.000000,0.720000)
(0.000000,0.700000)
(0.000000,0.680000)
(0.000000,0.660000)
(0.000000,0.640000)
(0.000000,0.620000)
(0.000000,0.600000)
(0.000000,0.580000)
(0.000000,0.560000)
(0.000000,0.540000)
(0.000000,0.520000)
(0.000000,0.500000)
(0.000000,0.480000)
(0.000000,0.460000)
(0.000000,0.440000)
(0.000000,0.420000)
(0.000000,0.420000)
(0.000000,0.380000)
(0.000000,0.360000)
(0.000000,0.340000)
(0.000000,0.320000)
(0.000000,0.300000)
(0.000000,0.280000)
(0.000000,0.260000)
(0.000000,0.240000)
(0.000000,0.220000)
(0.000000,0.200000)
(0.000000,0.180000)
(0.000000,0.160000)
(0.000000,0.140000)
(0.000000,0.120000)
(0.000000,0.100000)
(0.000000,0.080000)
(0.000000,0.060000)
(0.000000,0.040000)
(0.000000,0.020000)
(0.000000,0.000000)
};
\end{axis}
\end{tikzpicture}
\\
(a) Cridex & (b) Harebot
\\[2ex]
\begin{tikzpicture}[scale=0.50]
\begin{axis}[width=0.75\textwidth,height=0.675\textwidth,xmin=-0.02,xmax=1.0,
                   ymin=0.0,ymax=1.02,legend pos=south east,grid=both,
                   xlabel={False Positive Rate},ylabel={True Positive Rate}] 
\addplot[color=red,ultra thick] coordinates {
(1.000000,1.000000)
(1.000000,0.980000)
(0.980000,0.980000)
(0.960000,0.980000)
(0.940000,0.980000)
(0.920000,0.980000)
(0.900000,0.980000)
(0.880000,0.980000)
(0.860000,0.980000)
(0.840000,0.980000)
(0.820000,0.980000)
(0.800000,0.980000)
(0.780000,0.980000)
(0.760000,0.980000)
(0.740000,0.980000)
(0.720000,0.980000)
(0.700000,0.980000)
(0.680000,0.980000)
(0.660000,0.980000)
(0.640000,0.980000)
(0.620000,0.980000)
(0.600000,0.980000)
(0.580000,0.980000)
(0.560000,0.980000)
(0.540000,0.980000)
(0.520000,0.980000)
(0.500000,0.980000)
(0.480000,0.980000)
(0.460000,0.980000)
(0.440000,0.980000)
(0.420000,0.980000)
(0.400000,0.980000)
(0.380000,0.980000)
(0.360000,0.980000)
(0.340000,0.980000)
(0.320000,0.980000)
(0.300000,0.980000)
(0.280000,0.980000)
(0.260000,0.980000)
(0.240000,0.980000)
(0.220000,0.980000)
(0.200000,0.980000)
(0.180000,0.980000)
(0.160000,0.980000)
(0.140000,0.980000)
(0.120000,0.980000)
(0.100000,0.980000)
(0.080000,0.980000)
(0.060000,0.980000)
(0.040000,0.980000)
(0.020000,0.980000)
(0.000000,0.980000)
(0.000000,0.960000)
(0.000000,0.940000)
(0.000000,0.920000)
(0.000000,0.900000)
(0.000000,0.880000)
(0.000000,0.860000)
(0.000000,0.840000)
(0.000000,0.820000)
(0.000000,0.800000)
(0.000000,0.780000)
(0.000000,0.760000)
(0.000000,0.740000)
(0.000000,0.720000)
(0.000000,0.700000)
(0.000000,0.680000)
(0.000000,0.660000)
(0.000000,0.640000)
(0.000000,0.620000)
(0.000000,0.600000)
(0.000000,0.580000)
(0.000000,0.560000)
(0.000000,0.540000)
(0.000000,0.520000)
(0.000000,0.500000)
(0.000000,0.480000)
(0.000000,0.460000)
(0.000000,0.440000)
(0.000000,0.420000)
(0.000000,0.400000)
(0.000000,0.380000)
(0.000000,0.360000)
(0.000000,0.340000)
(0.000000,0.320000)
(0.000000,0.300000)
(0.000000,0.280000)
(0.000000,0.260000)
(0.000000,0.240000)
(0.000000,0.220000)
(0.000000,0.220000)
(0.000000,0.180000)
(0.000000,0.160000)
(0.000000,0.140000)
(0.000000,0.120000)
(0.000000,0.100000)
(0.000000,0.080000)
(0.000000,0.060000)
(0.000000,0.040000)
(0.000000,0.020000)
(0.000000,0.000000)
};
\end{axis}
\end{tikzpicture}
&
\begin{tikzpicture}[scale=0.50]
\begin{axis}[width=0.75\textwidth,height=0.675\textwidth,xmin=-0.02,xmax=1.0,
                   ymin=0.0,ymax=1.02,legend pos=south east,grid=both,
                   xlabel={False Positive Rate},ylabel={True Positive Rate}] 
\addplot[color=red,ultra thick] coordinates {
(1.000000,1.000000)
(0.990000,1.000000)
(0.980000,1.000000)
(0.970000,1.000000)
(0.960000,1.000000)
(0.950000,1.000000)
(0.940000,1.000000)
(0.930000,1.000000)
(0.920000,1.000000)
(0.910000,1.000000)
(0.900000,1.000000)
(0.890000,1.000000)
(0.880000,1.000000)
(0.870000,1.000000)
(0.860000,1.000000)
(0.850000,1.000000)
(0.840000,1.000000)
(0.830000,1.000000)
(0.820000,1.000000)
(0.810000,1.000000)
(0.800000,1.000000)
(0.790000,1.000000)
(0.780000,1.000000)
(0.770000,1.000000)
(0.760000,1.000000)
(0.750000,1.000000)
(0.740000,1.000000)
(0.730000,1.000000)
(0.720000,1.000000)
(0.710000,1.000000)
(0.700000,1.000000)
(0.690000,1.000000)
(0.680000,1.000000)
(0.670000,1.000000)
(0.660000,1.000000)
(0.650000,1.000000)
(0.640000,1.000000)
(0.630000,1.000000)
(0.620000,1.000000)
(0.610000,1.000000)
(0.600000,1.000000)
(0.590000,1.000000)
(0.580000,1.000000)
(0.570000,1.000000)
(0.560000,1.000000)
(0.550000,1.000000)
(0.540000,1.000000)
(0.530000,1.000000)
(0.520000,1.000000)
(0.510000,1.000000)
(0.500000,1.000000)
(0.500000,0.990000)
(0.490000,0.990000)
(0.480000,0.990000)
(0.470000,0.990000)
(0.460000,0.990000)
(0.450000,0.990000)
(0.440000,0.990000)
(0.430000,0.990000)
(0.420000,0.990000)
(0.410000,0.990000)
(0.400000,0.990000)
(0.390000,0.990000)
(0.380000,0.990000)
(0.370000,0.990000)
(0.360000,0.990000)
(0.350000,0.990000)
(0.340000,0.990000)
(0.330000,0.990000)
(0.320000,0.990000)
(0.310000,0.990000)
(0.300000,0.990000)
(0.290000,0.990000)
(0.280000,0.990000)
(0.270000,0.990000)
(0.260000,0.990000)
(0.250000,0.990000)
(0.240000,0.990000)
(0.240000,0.980000)
(0.230000,0.980000)
(0.220000,0.980000)
(0.210000,0.980000)
(0.200000,0.980000)
(0.190000,0.980000)
(0.180000,0.980000)
(0.170000,0.980000)
(0.160000,0.980000)
(0.150000,0.980000)
(0.140000,0.980000)
(0.130000,0.980000)
(0.130000,0.970000)
(0.120000,0.970000)
(0.110000,0.970000)
(0.110000,0.960000)
(0.110000,0.950000)
(0.100000,0.950000)
(0.090000,0.950000)
(0.090000,0.940000)
(0.080000,0.940000)
(0.070000,0.940000)
(0.070000,0.930000)
(0.070000,0.920000)
(0.060000,0.920000)
(0.060000,0.910000)
(0.060000,0.900000)
(0.050000,0.900000)
(0.040000,0.900000)
(0.030000,0.900000)
(0.020000,0.900000)
(0.010000,0.900000)
(0.000000,0.900000)
(0.000000,0.890000)
(0.000000,0.880000)
(0.000000,0.870000)
(0.000000,0.860000)
(0.000000,0.850000)
(0.000000,0.840000)
(0.000000,0.830000)
(0.000000,0.820000)
(0.000000,0.810000)
(0.000000,0.800000)
(0.000000,0.790000)
(0.000000,0.780000)
(0.000000,0.770000)
(0.000000,0.760000)
(0.000000,0.750000)
(0.000000,0.740000)
(0.000000,0.730000)
(0.000000,0.720000)
(0.000000,0.710000)
(0.000000,0.700000)
(0.000000,0.690000)
(0.000000,0.680000)
(0.000000,0.670000)
(0.000000,0.660000)
(0.000000,0.650000)
(0.000000,0.640000)
(0.000000,0.630000)
(0.000000,0.620000)
(0.000000,0.610000)
(0.000000,0.600000)
(0.000000,0.590000)
(0.000000,0.580000)
(0.000000,0.570000)
(0.000000,0.560000)
(0.000000,0.550000)
(0.000000,0.540000)
(0.000000,0.530000)
(0.000000,0.520000)
(0.000000,0.510000)
(0.000000,0.500000)
(0.000000,0.490000)
(0.000000,0.480000)
(0.000000,0.470000)
(0.000000,0.460000)
(0.000000,0.450000)
(0.000000,0.440000)
(0.000000,0.430000)
(0.000000,0.420000)
(0.000000,0.410000)
(0.000000,0.400000)
(0.000000,0.390000)
(0.000000,0.380000)
(0.000000,0.370000)
(0.000000,0.360000)
(0.000000,0.350000)
(0.000000,0.340000)
(0.000000,0.330000)
(0.000000,0.320000)
(0.000000,0.310000)
(0.000000,0.300000)
(0.000000,0.290000)
(0.000000,0.280000)
(0.000000,0.270000)
(0.000000,0.260000)
(0.000000,0.250000)
(0.000000,0.240000)
(0.000000,0.230000)
(0.000000,0.220000)
(0.000000,0.210000)
(0.000000,0.200000)
(0.000000,0.190000)
(0.000000,0.180000)
(0.000000,0.170000)
(0.000000,0.160000)
(0.000000,0.150000)
(0.000000,0.140000)
(0.000000,0.130000)
(0.000000,0.120000)
(0.000000,0.110000)
(0.000000,0.100000)
(0.000000,0.090000)
(0.000000,0.080000)
(0.000000,0.070000)
(0.000000,0.060000)
(0.000000,0.050000)
(0.000000,0.040000)
(0.000000,0.030000)
(0.000000,0.020000)
(0.000000,0.010000)
(0.000000,0.000000)
};
\end{axis}
\end{tikzpicture}
\\
(c) Smart HDD & (d) Winwebsec
\\[2ex]
\begin{tikzpicture}[scale=0.50]
\begin{axis}[width=0.75\textwidth,height=0.675\textwidth,xmin=-0.02,xmax=1.0,
                   ymin=0.0,ymax=1.02,legend pos=south east,grid=both,
                   xlabel={False Positive Rate},ylabel={True Positive Rate}] 
\addplot[color=red,ultra thick] coordinates {
(1.000000,1.000000)
(0.990000,1.000000)
(0.980000,1.000000)
(0.970000,1.000000)
(0.960000,1.000000)
(0.950000,1.000000)
(0.940000,1.000000)
(0.930000,1.000000)
(0.920000,1.000000)
(0.910000,1.000000)
(0.900000,1.000000)
(0.890000,1.000000)
(0.880000,1.000000)
(0.870000,1.000000)
(0.860000,1.000000)
(0.850000,1.000000)
(0.840000,1.000000)
(0.830000,1.000000)
(0.820000,1.000000)
(0.810000,1.000000)
(0.800000,1.000000)
(0.790000,1.000000)
(0.780000,1.000000)
(0.770000,1.000000)
(0.760000,1.000000)
(0.750000,1.000000)
(0.740000,1.000000)
(0.740000,0.990000)
(0.730000,0.990000)
(0.720000,0.990000)
(0.710000,0.990000)
(0.700000,0.990000)
(0.690000,0.990000)
(0.680000,0.990000)
(0.670000,0.990000)
(0.660000,0.990000)
(0.650000,0.990000)
(0.640000,0.990000)
(0.630000,0.990000)
(0.620000,0.990000)
(0.610000,0.990000)
(0.600000,0.990000)
(0.590000,0.990000)
(0.580000,0.990000)
(0.570000,0.990000)
(0.560000,0.990000)
(0.550000,0.990000)
(0.540000,0.990000)
(0.530000,0.990000)
(0.520000,0.990000)
(0.510000,0.990000)
(0.500000,0.990000)
(0.490000,0.990000)
(0.480000,0.990000)
(0.470000,0.990000)
(0.460000,0.990000)
(0.450000,0.990000)
(0.440000,0.990000)
(0.430000,0.990000)
(0.420000,0.990000)
(0.410000,0.990000)
(0.400000,0.990000)
(0.390000,0.990000)
(0.380000,0.990000)
(0.370000,0.990000)
(0.360000,0.990000)
(0.350000,0.990000)
(0.340000,0.990000)
(0.330000,0.990000)
(0.320000,0.990000)
(0.310000,0.990000)
(0.300000,0.990000)
(0.290000,0.990000)
(0.280000,0.990000)
(0.270000,0.990000)
(0.260000,0.990000)
(0.250000,0.990000)
(0.240000,0.990000)
(0.230000,0.990000)
(0.220000,0.990000)
(0.210000,0.990000)
(0.200000,0.990000)
(0.190000,0.990000)
(0.190000,0.980000)
(0.180000,0.980000)
(0.170000,0.980000)
(0.160000,0.980000)
(0.150000,0.980000)
(0.140000,0.980000)
(0.130000,0.980000)
(0.120000,0.980000)
(0.110000,0.980000)
(0.100000,0.980000)
(0.090000,0.980000)
(0.080000,0.980000)
(0.070000,0.980000)
(0.060000,0.980000)
(0.050000,0.980000)
(0.040000,0.980000)
(0.030000,0.980000)
(0.020000,0.980000)
(0.010000,0.980000)
(0.000000,0.980000)
(0.000000,0.970000)
(0.000000,0.960000)
(0.000000,0.950000)
(0.000000,0.940000)
(0.000000,0.930000)
(0.000000,0.920000)
(0.000000,0.910000)
(0.000000,0.900000)
(0.000000,0.890000)
(0.000000,0.880000)
(0.000000,0.870000)
(0.000000,0.860000)
(0.000000,0.850000)
(0.000000,0.840000)
(0.000000,0.830000)
(0.000000,0.820000)
(0.000000,0.810000)
(0.000000,0.800000)
(0.000000,0.790000)
(0.000000,0.780000)
(0.000000,0.770000)
(0.000000,0.760000)
(0.000000,0.750000)
(0.000000,0.740000)
(0.000000,0.730000)
(0.000000,0.720000)
(0.000000,0.710000)
(0.000000,0.700000)
(0.000000,0.690000)
(0.000000,0.680000)
(0.000000,0.670000)
(0.000000,0.660000)
(0.000000,0.650000)
(0.000000,0.640000)
(0.000000,0.630000)
(0.000000,0.620000)
(0.000000,0.610000)
(0.000000,0.600000)
(0.000000,0.590000)
(0.000000,0.580000)
(0.000000,0.570000)
(0.000000,0.560000)
(0.000000,0.550000)
(0.000000,0.540000)
(0.000000,0.530000)
(0.000000,0.520000)
(0.000000,0.510000)
(0.000000,0.500000)
(0.000000,0.490000)
(0.000000,0.480000)
(0.000000,0.470000)
(0.000000,0.460000)
(0.000000,0.450000)
(0.000000,0.440000)
(0.000000,0.430000)
(0.000000,0.420000)
(0.000000,0.410000)
(0.000000,0.400000)
(0.000000,0.390000)
(0.000000,0.380000)
(0.000000,0.370000)
(0.000000,0.360000)
(0.000000,0.350000)
(0.000000,0.340000)
(0.000000,0.330000)
(0.000000,0.320000)
(0.000000,0.310000)
(0.000000,0.300000)
(0.000000,0.290000)
(0.000000,0.280000)
(0.000000,0.270000)
(0.000000,0.260000)
(0.000000,0.250000)
(0.000000,0.240000)
(0.000000,0.230000)
(0.000000,0.220000)
(0.000000,0.210000)
(0.000000,0.200000)
(0.000000,0.190000)
(0.000000,0.180000)
(0.000000,0.170000)
(0.000000,0.160000)
(0.000000,0.150000)
(0.000000,0.140000)
(0.000000,0.130000)
(0.000000,0.120000)
(0.000000,0.110000)
(0.000000,0.100000)
(0.000000,0.090000)
(0.000000,0.080000)
(0.000000,0.070000)
(0.000000,0.060000)
(0.000000,0.050000)
(0.000000,0.040000)
(0.000000,0.030000)
(0.000000,0.020000)
(0.000000,0.010000)
(0.000000,0.000000)
};
\end{axis}
\end{tikzpicture}
&
\begin{tikzpicture}[scale=0.50]
\begin{axis}[width=0.75\textwidth,height=0.675\textwidth,xmin=-0.02,xmax=1.0,
                   ymin=0.0,ymax=1.02,legend pos=south east,grid=both,
                   xlabel={False Positive Rate},ylabel={True Positive Rate}] 
\addplot[color=red,ultra thick] coordinates {
(1.000000,1.000000)
(0.990000,1.000000)
(0.990000,0.990000)
(0.980000,0.990000)
(0.970000,0.990000)
(0.960000,0.990000)
(0.950000,0.990000)
(0.940000,0.990000)
(0.930000,0.990000)
(0.920000,0.990000)
(0.920000,0.980000)
(0.910000,0.980000)
(0.900000,0.980000)
(0.890000,0.980000)
(0.880000,0.980000)
(0.870000,0.980000)
(0.860000,0.980000)
(0.850000,0.980000)
(0.840000,0.980000)
(0.830000,0.980000)
(0.820000,0.980000)
(0.810000,0.980000)
(0.800000,0.980000)
(0.790000,0.980000)
(0.780000,0.980000)
(0.770000,0.980000)
(0.760000,0.980000)
(0.750000,0.980000)
(0.740000,0.980000)
(0.730000,0.980000)
(0.720000,0.980000)
(0.710000,0.980000)
(0.700000,0.980000)
(0.690000,0.980000)
(0.680000,0.980000)
(0.670000,0.980000)
(0.660000,0.980000)
(0.650000,0.980000)
(0.640000,0.980000)
(0.630000,0.980000)
(0.620000,0.980000)
(0.610000,0.980000)
(0.600000,0.980000)
(0.590000,0.980000)
(0.580000,0.980000)
(0.570000,0.980000)
(0.560000,0.980000)
(0.550000,0.980000)
(0.540000,0.980000)
(0.530000,0.980000)
(0.520000,0.980000)
(0.510000,0.980000)
(0.500000,0.980000)
(0.490000,0.980000)
(0.480000,0.980000)
(0.470000,0.980000)
(0.460000,0.980000)
(0.450000,0.980000)
(0.440000,0.980000)
(0.430000,0.980000)
(0.420000,0.980000)
(0.410000,0.980000)
(0.400000,0.980000)
(0.390000,0.980000)
(0.380000,0.980000)
(0.370000,0.980000)
(0.360000,0.980000)
(0.350000,0.980000)
(0.340000,0.980000)
(0.330000,0.980000)
(0.320000,0.980000)
(0.310000,0.980000)
(0.300000,0.980000)
(0.290000,0.980000)
(0.280000,0.980000)
(0.270000,0.980000)
(0.260000,0.980000)
(0.250000,0.980000)
(0.240000,0.980000)
(0.230000,0.980000)
(0.220000,0.980000)
(0.210000,0.980000)
(0.200000,0.980000)
(0.190000,0.980000)
(0.180000,0.980000)
(0.170000,0.980000)
(0.160000,0.980000)
(0.150000,0.980000)
(0.140000,0.980000)
(0.130000,0.980000)
(0.120000,0.980000)
(0.110000,0.980000)
(0.100000,0.980000)
(0.090000,0.980000)
(0.080000,0.980000)
(0.070000,0.980000)
(0.060000,0.980000)
(0.050000,0.980000)
(0.040000,0.980000)
(0.040000,0.970000)
(0.030000,0.970000)
(0.020000,0.970000)
(0.010000,0.970000)
(0.010000,0.960000)
(0.010000,0.950000)
(0.010000,0.940000)
(0.010000,0.930000)
(0.010000,0.920000)
(0.010000,0.910000)
(0.010000,0.900000)
(0.000000,0.900000)
(0.000000,0.890000)
(0.000000,0.890000)
(0.000000,0.870000)
(0.000000,0.860000)
(0.000000,0.850000)
(0.000000,0.840000)
(0.000000,0.830000)
(0.000000,0.830000)
(0.000000,0.810000)
(0.000000,0.800000)
(0.000000,0.790000)
(0.000000,0.780000)
(0.000000,0.770000)
(0.000000,0.760000)
(0.000000,0.750000)
(0.000000,0.740000)
(0.000000,0.730000)
(0.000000,0.720000)
(0.000000,0.710000)
(0.000000,0.700000)
(0.000000,0.690000)
(0.000000,0.680000)
(0.000000,0.680000)
(0.000000,0.660000)
(0.000000,0.650000)
(0.000000,0.640000)
(0.000000,0.630000)
(0.000000,0.630000)
(0.000000,0.610000)
(0.000000,0.600000)
(0.000000,0.600000)
(0.000000,0.580000)
(0.000000,0.570000)
(0.000000,0.560000)
(0.000000,0.550000)
(0.000000,0.550000)
(0.000000,0.530000)
(0.000000,0.520000)
(0.000000,0.510000)
(0.000000,0.500000)
(0.000000,0.490000)
(0.000000,0.480000)
(0.000000,0.480000)
(0.000000,0.460000)
(0.000000,0.450000)
(0.000000,0.440000)
(0.000000,0.440000)
(0.000000,0.440000)
(0.000000,0.410000)
(0.000000,0.400000)
(0.000000,0.400000)
(0.000000,0.380000)
(0.000000,0.380000)
(0.000000,0.380000)
(0.000000,0.350000)
(0.000000,0.340000)
(0.000000,0.330000)
(0.000000,0.320000)
(0.000000,0.310000)
(0.000000,0.300000)
(0.000000,0.290000)
(0.000000,0.280000)
(0.000000,0.270000)
(0.000000,0.260000)
(0.000000,0.250000)
(0.000000,0.240000)
(0.000000,0.230000)
(0.000000,0.220000)
(0.000000,0.220000)
(0.000000,0.200000)
(0.000000,0.190000)
(0.000000,0.190000)
(0.000000,0.170000)
(0.000000,0.160000)
(0.000000,0.150000)
(0.000000,0.150000)
(0.000000,0.130000)
(0.000000,0.120000)
(0.000000,0.110000)
(0.000000,0.100000)
(0.000000,0.100000)
(0.000000,0.100000)
(0.000000,0.100000)
(0.000000,0.060000)
(0.000000,0.050000)
(0.000000,0.040000)
(0.000000,0.030000)
(0.000000,0.020000)
(0.000000,0.010000)
(0.000000,0.000000)
};
\end{axis}
\end{tikzpicture}
\\
(e) Zbot & (f) Zeroaccess
\\[2ex]
\end{tabular}
\end{center}
\vglue -0.20in
\caption{ROC Curves for HMMs Based on Dynamic Birthmarks\label{fig:d_roc}}
\end{figure*}

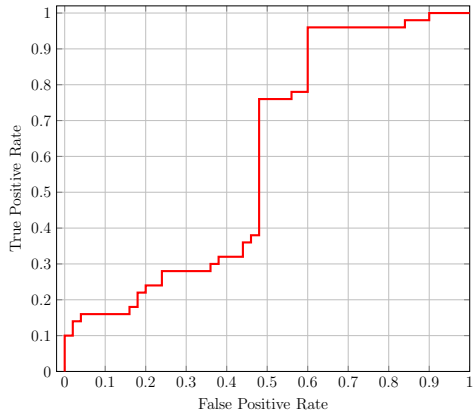
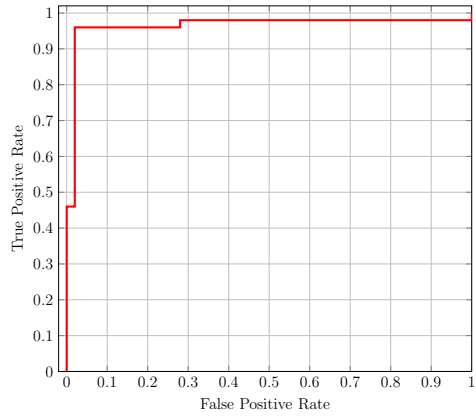
\end{document}